\documentclass[review,3p,sort&compress]{elsarticle}
\usepackage{hyperref}
\usepackage{amsmath,amssymb}
\usepackage[noend]{algpseudocode}
\usepackage{algorithm}
\usepackage{bm}
\usepackage{graphicx}
\usepackage{subcaption}
\usepackage{pgfplots}
\pgfplotsset{
compat=newest,
table/search path={results/},
colormap name=viridis,
max space between ticks=15,
}
\usetikzlibrary[pgfplots.groupplots]

\newcommand{\Rbb}{\mathbb{R}}
\newcommand{\Cbb}{\mathbb{C}}
\newcommand{\cmdl}{^{(i,\lambda)}}
\newcommand{\cmdi}{^{(i)}}
\newcommand{\cmdim}{^{(i-1)}}

\newcommand{\cmdhat}[1]{\hat{#1}}

\newcommand{\trans}{^\mathsf{T}}
\newcommand{\herm}{^\mathsf{H}}
\newcommand{\cmdt}{\bm{\theta}}
\newcommand{\cmdp}{\bm{\phi}}
\newcommand{\cmdP}{\bm{\Phi}}

\newcommand{\A}{\bm{A}}
\newcommand{\Ab}{\bar{\bm{A}}}
\newcommand{\ab}{\bar{\bm{a}}}
\newcommand{\at}{\hat{\bm{a}}}
\newcommand{\cmdA}{\A(\cmdt,\cmdP,n)}
\newcommand{\cmdxbar}{\bar{\bm{x}}}
\newcommand{\cmdr}{\bar{\bm{r}}}

\newcommand{\cmdAbar}{\Ab(\cmdt,\cmdP)}
\newcommand{\cmdAbarpsi}{\Ab}
\newcommand{\cmdAbarherm}{\Ab\herm(\cmdt,\cmdP)}
\newcommand{\cmdPseudo}{\Ab^\dagger(\cmdt,\cmdP)}
\newcommand{\cmdPi}{\bm{\Pi}}
\newcommand{\cmdorthPi}{\bm{\Pi}^\perp}
\newcommand{\cmdProj}{\bm{\Pi}_{\Ab}(\cmdt,\cmdP)}
\newcommand{\cmdorthproj}{\bm{\Pi}_{\Ab}^\perp(\cmdt,\cmdP)}

\newcommand{\cmddegree}[1]{{#1}^{\circ}}

\DeclareMathOperator*{\argmin}{arg\,min} \DeclareMathOperator*{\argmax}{arg\,max}  
\journal{Signal Processing}
\begin{document}
\begin{frontmatter}
\title{Sequential Maximum-Likelihood Estimation of Wideband Polynomial-Phase Signals on Sensor Array}
\author[1]{Kaleb Debre}
\ead{kaleb.debre@tu-darmstadt.de}
\author[2]{Tai Fei}
\ead{tai.fei@fh-dortmund.de}
\author[1]{Marius Pesavento}
\ead{marius.pesavento@tu-darmstadt.de}
\affiliation[1]{organization={Communication Systems Group, Technische Universität Darmstadt},city={64283 Darmstadt},country={ Germany}}
\affiliation[2]{organization={Department of Information Technology, Fachhochschule Dortmund},city={44139 Dortmund},country={ Germany}}
\begin{highlights}
\item FMCW radar detection of multiple targets with time-varying radial velocities.
\item  Joint estimation of direction-of-arrival and time-frequency signatures.
\item Computationally tractable approximate Maximum-likelihood estimation method.
\item  Wideband space-time-frequency coherent processing
\item Achieves Cramér-Rao bound for all parameters in challenging multi-source scenarios.
\end{highlights}
\begin{abstract}
This paper presents a novel sequential estimator for the direction-of-arrival and polynomial coefficients of wideband polynomial-phase signals impinging on a sensor array.
Addressing the computational challenges of Maximum-likelihood estimation for this problem, we propose a method leveraging random sampling consensus (RANSAC) applied to the time-frequency spatial signatures of sources. Our approach supports multiple sources and higher-order polynomials by employing coherent array processing and sequential approximations of the Maximum-likelihood cost function. We also propose a low-complexity variant that estimates source directions via angular domain random sampling.
Numerical evaluations demonstrate that the proposed methods achieve Cramér-Rao bounds in challenging multi-source scenarios, including closely spaced time-frequency spatial signatures, highlighting their suitability for advanced radar signal processing applications.\end{abstract}
\end{frontmatter}

\section{Introduction}
Polynomial-phase signals (PPS) emerge in several important applications including radar, short-range sonar and ultra-sound imaging, where frequency-modulated continuous-wave (FMCW) signals are commonly used \cite{stankovic2014a,blunt2018,levanon2004}.
Conventional FMCW radar receivers employ stretch processing, where the received signal is mixed with the transmitted linear chirp signal to produce a low-frequency beat signal for processing \cite{al-hourani2018,richards2014}.
While this approach benefits from lower sampling rate requirements, it can be sensitive to phase noise in the local oscillator, which is more pronounced as the range increases due to decorrelation of the received and transmitted signal, leading to masking effects that obscure weaker targets in the far range \cite{skolnik2008}.
Traditional Doppler processing is carried out in Doppler bins based on the assumption of constant relative velocity between targets and transceivers inducing constant frequency shifts (linear phase shift) in the radar return signal \cite{richards2014}.
However, when the relative velocity is varying due to acceleration and jerk of targets, the Doppler shift is more precisely characterized by a higher-order polynomial-phase shift \cite{kelly1961,rihaczek1996}. 
Such effects can also arise due to changes in the relative angle between targets and transceivers, which is especially pronounced in near- and mid-field radar applications \cite{skolnik1980}.
Time-varying Doppler also impacts medical ultrasound imaging and blood velocity estimation, where the slower propagation speed of sound produces significant Doppler effects \cite{luijten2023,gudmundson2011}.

Recently, software-defined radio architectures have been proposed, where the received signal is directly sampled using high-performance analog-to-digital converters with high sampling rates and the signal processing is performed entirely in software, eliminating the need for analog mixers.
These direct-sampling architectures are particularly advantageous for enabling flexible radar receivers, including scenarios that involve non-linear target motions and spectrum congestion \cite{ohagan2018,debatty2010}.
Information about the target's dynamics and trajectory, characterized by its time-varying Doppler signature, is encoded in the polynomial-phase coefficients of the return signal \cite{peleg1995}. 
Furthermore, angular information is available from the phase differences of the return signal impinging on different antennas.
Therefore, modern radar systems that utilize direct sampling of received signals offer great opportunities in radar signal processing for target identification and tracking by designing appropriate parameter estimation algorithms.

The problem of estimating the polynomial coefficients of a PPS has thus drawn much attention in the literature.
The corresponding Maximum-likelihood (ML) estimator has high computational cost for polynomial orders greater than one \cite{peleg1996}. 
Low-complexity PPS estimation using the high-order ambiguity function (HAF) has been studied extensively \cite{djuric1990,peleg1991a,peleg1995,djurovic2012b,barbarossa1998,pham2007}.
Recently, the random sampling consensus (RANSAC) method, which had originally been proposed in the context of computer vision \cite{fischler1981}, has been used for parameter estimation of multiple PPSs by sub-sampling instantaneous frequencies in the time-frequency domain \cite{djurovic2017,djurovic2018}. 
When considering PPSs impinging on a sensor array, the corresponding problem is wideband direction-of-arrival (DOA) estimation \cite{viberg2014,amin2014}.
It has been demonstrated in \cite{ma2006} that, when the deterministic signal structure is accounted in the signal model, coherent processing of the array measurements can significantly improve the DOA estimation performance, compared to a stochastic signal model. 

The optimal ML estimator for jointly estimating the DOA and polynomial coefficients of multiple wideband PPSs impinging on a uniform linear array has been proposed in \cite{gershman2001}.
This joint DOA-PPS estimator has high computational cost due to the required exhaustive search in the multi-dimensional, multi-source parameter space.
To reduce this high complexity, the work in \cite{gershman2001} further introduces a DOA-PPS estimator based on the single-source approximation of the ML cost function, referred to as the polynomial-phase beamformer.
However, for high polynomial orders of the PPSs, the computational complexity of the polynomial-phase beamformer can still be prohibitive.
Therefore, recent DOA-PPS estimation techniques combine PPS estimation techniques, such as HAF-based methods, with a spectral search across the field of view using the polynomial-phase beamformer \cite{rakovic2017,djurovic2012a,djurovic2019}.
While these approaches demonstrate adequate performance in simple scenarios, they encounter significant challenges in handling multiple closely spaced sources, higher-order polynomials, and time-varying Doppler effects.
This is primarily due to the inability of the polynomial-phase beamformer to effectively address interference among multiple sources.
These challenges underscore the need for more robust methods that can reliably operate in complex, multi-source environments.

Our paper addresses these challenges through a novel sequential DOA-PPS estimation approach.
We leverage the inherent sparsity in the time-frequency spatial signatures of the sources using sensor-wise time-frequency analysis and coherent array processing of the measurements.
Following the approach of \cite{djurovic2017} for PPS estimation, we apply the RANSAC method to generate candidate DOA-PPS parameter estimates of one source at a time by sub-sampling time-frequency spatial source signatures.
To select the best-fitting candidate parameter estimates with the measurement data, we employ sequential approximations of the ML cost function and gradient-based parameter refinement to enhance accuracy.
Additionally, we present a low-cost variant, where source directions are estimated using angular domain random sampling and estimated sources are efficiently eliminated in the measurement data, further reducing the computational complexity.

Our method contributes to DOA-PPS estimation in several ways:
\begin{itemize}
	\item  We address the high computational cost of the ML estimator by estimating the parameters of one source at a time using the RANSAC method on the time-frequency spatial source signatures.
	\item Unlike existing methods, our approach is effective for multiple sources, including closely spaced, higher-order PPSs due to coherent array processing and sequential approximations of the ML cost function.
	\item We present a low-cost sequential estimator variant, where random sub-sampling is also applied in the angular domain and estimated sources are efficiently eliminated in the data.\item The proposed methods achieve the Cramér-Rao bound for all parameters in challenging multi-source scenarios, including scenarios with closely spaced time-frequency spatial source signatures.
\end{itemize}
The remainder of this paper is structured as follows.
Section \ref{spatiotemporalsignalmodel} defines the signal model.
Section \ref{MLestimator} reviews the ML-based DOA-PPS estimators.
Section \ref{proposed} presents our proposed sequential DOA-PPS estimator and its low-cost variant.
Numerical results are discussed in Section \ref{numericalresults}, followed by conclusions in Section \ref{conclusion}.

\textit{Notation:} Matrices and vectors are denoted by bold-faced, capital and lower-case letters, respectively.
The superscripts $(\cdot)^*$, $(\cdot)^\mathsf{T}$ and $(\cdot)^\mathsf{H}$ denote complex-conjugate, transpose and Hermitian transpose, respectively.
The subscripts $[\cdot]_{j}$, $[\cdot]_{:,j}$ and $[\cdot]_{i,j}$ denote the $j$th element of a vector, the $j$th column of a matrix and the element of a matrix in its $i$th row and $j$th column, respectively.
The operators $\|\cdot\|_2$ and $\mathcal{O}(\cdot)$  denote the $\ell_2$-norm of a vector and the Big-O notation, respectively.
The operator $|\cdot|$ denotes both the absolute value and the cardinality of a set.
The operators $\lceil\cdot\rceil$ and $\lfloor\cdot\rfloor$ denote rounding up and rounding down to the nearest integer value, respectively.
The symbol $\bm{I}_M$ denotes the identity matrix of dimension $M \times M$.

\section{Signal Model}\label{spatiotemporalsignalmodel}
Consider $L$ far-field, passband polynomial-phase signals of polynomial order $K$.
The $\ell$th PPS for $\ell=1,\ldots,L$ is defined as
\begin{equation}
	s_\ell(t)=\alpha_\ell e^{\mathrm{j}\sum_{k=1}^{K}\phi_{\ell,k}t^k},\label{transmit}
\end{equation}
where, without loss of generality, it is assumed that the initial phases $\phi_{\ell,0}$ are absorbed in the signal amplitude $\alpha_\ell\in\Cbb$ for $\ell=1,\ldots,L$. The polynomial coefficients of the $\ell$th PPS are contained in the vector $\cmdp_\ell=\left[\phi_{\ell,1},\phi_{\ell,2},\ldots,\phi_{\ell,K}\right]\trans\in\Rbb^K$ and the physical unit of the $k$th-order polynomial coefficient $\phi_{\ell,k}$ is $\frac{\text{rad}}{\text{s}^k}$ for $k=1,\ldots,K$.
The signal phase of the $\ell$th PPS and its continuous-time derivative, referred to as the instantaneous frequency, are defined as
\begin{align}\label{phasefrequency}
	\varphi_\ell(t)&=\sum_{k=1}^{K}{\phi}_{\ell,k} t^k,\\
	\omega_\ell(t)&=\frac{d \varphi_\ell(t)}{dt}=\sum_{k=1}^{K}k{\phi}_{\ell,k} t^{k-1}.
\end{align}
The superposition of the $L$ PPSs impinges on a uniform linear array consisting of $M$ omnidirectional sensors, where 
it is sampled at discrete times $t=\Delta \cdot n$ for $n=-\left\lceil\frac{N-1}{2}\right\rceil,\ldots,\left\lfloor\frac{N-1}{2}\right\rfloor$ \cite{ristic1998,peleg1991}.
The symbols $\Delta$ and $N$ denote the sampling period in seconds and number of available snapshots, respectively. 

The corresponding baseband received signal vector ${\bm{x}(n)=\left[x_1(n),\ldots,x_{M}(n)\right]}\trans\in\Cbb^M$ is consequently given by
\begin{equation}
	\bm{x}(n)=\cmdA\bm{\alpha}+\bm{z}(n)\label{received2},
\end{equation}
where $\bm{z}(n)\in\Cbb^M$ denotes spatially and temporally white complex circular Gaussian sensor noise with noise power $\sigma^2$.
The vectors ${\bm{\alpha}=\left[\alpha_1,\ldots,\alpha_L\right]\trans\in\Cbb^L}$,  $\cmdt=\left[\theta_1,  \ldots, \theta_L\right]\trans\in\Rbb^L$ and matrix $\cmdP=\left[\cmdp_1,\ldots,\cmdp_L\right]\in\Rbb^{K\times L}$ summarize the unknown signal amplitudes, DOAs relative to broadside and polynomial-phase coefficients of the $L$ PPS sources, respectively. 
The elements of the time-varying array response matrix $\cmdA \in \Cbb^{M \times L}$ are defined as
\begin{equation}
	{\left[\cmdA\right]_{m,\ell}=e^{\mathrm{j}\sum_{k=1}^{K}{\phi}_{\ell,k} (\Delta n-(m-1)\tau_\ell)^k}},\label{wideband}
\end{equation}
where $m=1,\ldots,M$ denotes the sensor index.
The time delay in seconds between the arrival of the wave on adjacent sensors is given by \begin{equation}
	\tau=\frac{d}{c}\sin\vartheta\label{isd},
\end{equation}
where $\vartheta\in[-\frac{\pi}{2},\frac{\pi}{2}]$ denotes the source angle in the field-of-view, hence, $\tau_\ell=\frac{d}{c}\sin\theta_\ell$ for $\ell=1,\ldots,L$.
The constants $d$ and $c$ denote the distance between adjacent sensors and the propagation speed of the wave, respectively.
It is worth noting that the array response matrix $\cmdA$ in \eqref{wideband} incorporates both the temporal phase component of the transmitted signal in \eqref{transmit} and the spatial phase component corresponding to the wavefront arrival at the sensors.
The $\ell$th column of the array response matrix, denoted as $\bm{a}(\theta_\ell,\bm{\phi}_\ell,n)=\left[\cmdA\right]_{:,\ell}$, represents the array response vector of the $\ell$th source and is solely determined by its DOA $\theta_\ell$ and PPS parameters $\cmdp_\ell$.

\section{Maximum-Likelihood-Based DOA-PPS Estimation\label{MLestimator}}
The ML estimator for the multi-source DOA-PPS estimation problem and its single-source approximation, referred to as the polynomial-phase beamformer, have been proposed in \cite{gershman2001}.
The received signal vector in \eqref{received2} can be compactly written as \begin{equation}
	\bar{\bm{x}}=\cmdAbar\bm{\alpha}+\bar{\bm{z}},\label{stacked}
\end{equation}
where the received signal vectors $\bm{x}(n)$ for all $N$ snapshots are stacked in a tall vector as $\bar{\bm{x}} = [{\bm{x}}(-\lceil\frac{N-1}{2}\rceil)\trans, \ldots,\\
{\bm{x}}(\left\lfloor\frac{N-1}{2}\right\rfloor)^{\mathsf{T}}]\trans\in\Cbb^{MN}$.
The tall array response matrix and noise vector are accordingly given by $\cmdAbar = [\A(\cmdt,\cmdP,-\left\lceil\frac{N-1}{2}\right\rceil)\trans ,\ldots,\A(\cmdt,\cmdP,\left\lfloor\frac{N-1}{2}\right\rfloor)\trans ]\trans\in\Cbb^{MN\times L}$ and $\bar{\bm{z}} = [{\bm{z}}(-\left\lceil\frac{N-1}{2}\right\rceil)\trans, \ldots, {\bm{z}}(\left\lfloor\frac{N-1}{2}\right\rfloor)^{\mathsf{T}}]\trans\in\Cbb^{MN}$, respectively.

Omitting constant terms, the negative log-likelihood function corresponding to the signal model in \eqref{stacked} is given by
\begin{equation}
	f_L(\cmdt,\cmdP,\bm{\alpha})=\big\|\bar{\bm{x}}-\cmdAbar\bm{\alpha}\big\|_2^2\label{leastsquares}.
\end{equation}
The closed-form minimizer of \eqref{leastsquares} w.r.t. the signal amplitude vector $\bm{\alpha}$ yields 
\begin{equation}\label{concentration}
\hat{\bm{\alpha}} = \cmdPseudo \bar{\bm{x}},
\end{equation}
where ${\cmdPseudo=\left(\cmdAbarherm\cmdAbar\right)^{-1}\cmdAbarherm\in\Cbb^{L\times MN}}$ denotes the Moore-Penrose pseudo-inverse.
Inserting \eqref{concentration} into \eqref{leastsquares} yields the concentrated negative log-likelihood function
\begin{equation}\label{likelihood}
	\begin{split}
		f_L(\cmdt,\cmdP)&=\big\|\bar{\bm{x}}-\cmdProj\cmdxbar\big\|_2^2,\\
		&=\big\|\cmdorthproj\bar{\bm{x}}\big\|_2^2,
	\end{split}
\end{equation}
where ${\cmdProj= \cmdAbar\cmdPseudo\in\Cbb^{MN\times MN}}$ and ${\cmdorthproj=\bm{I}_{MN}-\cmdProj}$ denote the projector onto the range space of $\cmdAbar$ and its orthogonal complement, respectively.

The optimal ML-based DOA-PPS estimator is then obtained as \cite{gershman2001}
\begin{equation}
	\left\{\cmdhat{\cmdt},\cmdhat{\cmdP}\right\}=\argmin_{{\cmdt\in\Rbb^{L},
			\cmdP\in\Rbb^{K\times L}}}	f_L(\cmdt,\cmdP)\label{jointMLestimator}.
\end{equation}
The ML cost function $f_L(\cmdt,\cmdP)$ is highly non-convex and multi-modal, hence, it is difficult to apply gradient-based search techniques to find the global minimum of \eqref{jointMLestimator}. The brute-force search method, which involves evaluating \eqref{likelihood} across the entire parameter space to find the global minimum, typically exhibits prohibitive computational complexity of order $\mathcal{O}(G^{L(K+1)} (MN)^2)$, where $G$ represents the number of grid points per dimension used in the uniform discretization of the source parameters.

To reduce the computational complexity of the ML estimator, an alternative DOA-PPS estimator was proposed in \cite{gershman2001} based on a single-source approximation, where the multi-source criterion in \eqref{likelihood} is simplified by assuming the presence of only one PPS source at a time.
The resulting sub-optimal DOA-PPS estimator is given by
\begin{equation}
	\left\{\cmdhat{\cmdt},\cmdhat{\cmdP}\right\}= {^L\hspace{-0.15cm}\argmax_{{\theta}\in\Rbb,\cmdp\in\Rbb^{K}}}	f_1(\theta,\cmdp).\label{SSAMLestimator}
\end{equation}
where $^L\hspace{-0.07cm}\argmax$ denotes the $L$ largest peaks in the search space. 
The corresponding cost function based on the single-source approximation, referred to as the polynomial-phase beamformer, is given by
\begin{align}\label{polynomialbeamformer}
f_1(\theta,\cmdp)&=\frac{1}{MN}\big|\bar{\bm{a}}^\mathsf{H}({\theta},\cmdp)\bar{\bm{x}}\big|^2,
	\end{align}
where $\bar{\bm{a}}({\theta},\cmdp)={[\cmdAbar]_{:,1}\in\Cbb^{MN}}$ denotes the array response vector of a single source with DOA $\theta$ and polynomial-phase coefficients $\cmdp=\left[\phi_{1},\phi_{2},\ldots,\phi_{K}\right]\trans\in\Rbb^K$. The polynomial-phase beamformer \eqref{polynomialbeamformer} does not account for inter-source interference, resulting in sub-optimal estimates from \eqref{SSAMLestimator}. Moreover, for high-order polynomial phase signals, the brute-force search in \eqref{SSAMLestimator} still exhibits prohibitive computational complexity of order $\mathcal{O}(G^{(K+1)}MN)$. These limitations emphasize the need for an efficient parameter estimation method capable of handling the high-dimensional parameter space of the ML cost function.

\section{Sequential DOA-PPS Estimator}\label{proposed}
This work proposes a sequential DOA-PPS estimation procedure, which addresses the high complexity of the ML estimator in \eqref{jointMLestimator}.
In each iteration, the DOA and PPS parameters of one source at a time are estimated.
Our approach leverages the inherent sparsity in the time-frequency spatial source signatures in the time-frequency beamspace representation of the measurement data.
The RANSAC method is applied to this sparse support to generate candidate estimates for the DOA and PPS parameters of one source at a time.
Each iteration also includes gradient-based parameter refinement to enhance estimation accuracy.

The best-fitting candidate parameter estimates are selected by evaluating their agreement with the measurement data using the $i$-source approximation of the ML cost function in \eqref{likelihood}.
This approximated cost function, denoted as the $i$-source ML approximation, is given by
\begin{align}\label{isource}
	f_i(\cmdt^{(i)},\cmdP^{(i)})&=\big\|\bm{\Pi}_{\bar {\bm A}}^\perp(\bm{\theta}^{(i)},\cmdP^{(i)})\bar{\bm{x}}\big\|_2^2,
\end{align}
where $i=1,\ldots,L$ denotes the $i$th iteration of the sequential estimation procedure, and $\{\bm{\theta}\cmdi$,$\cmdP\cmdi\}$ denotes the DOA and PPS parameters of the $i$ sources in the $i$th iteration.
Furthermore, the residual received signal vector is introduced, which is obtained in the $i$th sequential iteration by removing the contributions of the $i$ previously estimated sources from the original measurements, expressed as
\begin{equation}\label{residual}
	\cmdr\cmdi=\bm{\Pi}_{\bar {\A}}^\perp(\hat{\bm{\theta}}\cmdi,\hat{\cmdP}\cmdi)\cmdxbar,
\end{equation}
where $\cmdr^{(0)}=\cmdxbar$, and $\{\hat{\bm{\theta}}\cmdi$,$\hat{\cmdP}\cmdi\}$ represents the estimated DOA and PPS parameters of the $i$ estimated sources.

\subsection{Coherent Array Processing With PPSs\label{sec:dftbeamformers}}
The measurement data are analyzed in the time-frequency-beamspace domain to leverage the sparsity inherent in the time-frequency spatial signatures of the sources.
To avoid cross-terms in the case of multiple PPSs, the short-time Fourier transform (STFT) is used as a linear time-frequency spectrum estimator, rather than the non-linear Wigner-Ville distribution \cite{stankovic2014}.
In the $i$th iteration, the STFT in the $m$th sensor for $m=1,\ldots,M-1$ is formed from the unstacked residual received signal vector $r_m(n)=[\cmdr\cmdim]_{M(n+\left\lceil\frac{N-1}{2}\right\rceil)+m}$, obtained from $\cmdr\cmdim$ in \eqref{residual}, as
\begin{equation}\label{y_m}
	y_m(p,q)=\frac{1}{H}\hspace{-0.1cm}\sum_{h=-\left\lceil\frac{H-1}{2}\right\rceil}^{\left\lfloor\frac{H-1}{2}\right\rfloor}\hspace{-0.1cm}r_m(p+h)e^{-\mathrm{j}\frac{2\pi h}{H} q},\end{equation}
where $p$ and $q$ denote the time and frequency index, respectively, and the rectangular window function of width $H$ is used.

A bank of $M$ beamformers is further applied to the time-frequency spectra in \eqref{y_m} stacked along the sensor dimension, i.e., $\bm{y}(p,q)=\left[y_1(p,q),\ldots,y_M(p,q)\right]^\mathsf{T}\in\Cbb^M$. The normalized beamformer output for the $b$th beamformer for $b = 0,...,M-1$ is given by
\begin{equation}\label{z_b}
{z}_b(p,q)=\frac{1}{M}\bm{w}_b\herm\bm{y}(p,q),\end{equation}
where $\bm{w}_b\in\Cbb^M$ represents the beamformer weight vector \cite{vorobyov2014}.

Fully coherent processing of the received signal in \eqref{received2} requires time-variant beamformers that are matched to the different instantaneous frequencies of the PPSs.
However, to keep the processing computationally tractable, we consider in \eqref{z_b} time-invariant beamforming, which can only partially provide coherent processing, depending on the time-frequency spatial signatures of the signals.
We employ a bank of $M$ time-invariant discrete Fourier transform (DFT) beamformers as $[\bm{w}_b]_{m}=e^{\mathrm{j}\frac{2\pi (m-1)b}{M}}$, which, in the case of a uniform linear array, can be efficiently implemented using the fast Fourier transform \cite{vantrees2002}.  The normalized magnitude response of the $b$th DFT beamformer for a signal with instantaneous frequency $\omega$ impinging from direction $\vartheta\in[-\frac{\pi}{2},\frac{\pi}{2}]$, is given by \cite{viberg2014,vantrees2002}
\begin{align}\label{beampatterna}
	\begin{split}|P_b(\vartheta,\omega)| =\frac{1}{M}|\bm{w}_b\herm\bm{a}(\vartheta,\omega)|&=\frac{1}{M}\left|\sum_{m=1}^{M}e^{-\mathrm{j}(m-1)\left(\omega\frac{d}{c}\sin\vartheta+\frac{2\pi b}{M}\right)}\right|,\\
&=\frac{1}{M}\left|\frac{\sin\left(\frac{M}{2}\left(\omega\frac{d}{c}\sin\vartheta+\frac{2\pi b}{M}\right)\right)}{\sin\left(\frac{1}{2}\left(\omega\frac{d}{c}\sin\vartheta+\frac{2\pi b}{M}\right)\right)}\right|.
\end{split}
	\end{align}
The narrowband array response vector $\bm{a}(\vartheta,\omega)$ with $[\bm{a}(\vartheta,\omega)]_m=e^{-\mathrm{j}(m-1)\omega\frac{d}{c}\sin\vartheta}$ in \eqref{beampatterna} approximates the array response vector $\bm{a}(\vartheta,\cmdp,n)$ in \eqref{wideband} with $[\bm{a}(\vartheta,\cmdp,n)]_m=e^{\mathrm{j}\sum_{k=1}^{K}{\phi}_{k} (\Delta n-(m-1)\tau)^k}$. 
This approximation holds, if the instantaneous frequency $\omega(n)$ of the impinging signals remains approximately constant over short time intervals and across small groups of adjacent sensors. 
Under these conditions, ${\bm a}(\vartheta,\omega) \approx {\bm a}(\theta,\cmdp,n)$ is valid within a space-time interval $M_{\rm c, 1} \leq m \leq M_{\rm c,2}$ and $N_{\rm c,1}\leq n \leq N_{\rm c,2}$, where coherent array processing with PPSs can be achieved.
This coherent array processing interval can be specified based on the first-order Taylor approximation of the phase $\operatorname*{arg}([\bm{a}(\vartheta,\cmdp,n)]_m)=\sum_{k=1}^{K}{\phi}_{k} (\Delta n-(m-1)\tau)^k$ in \eqref{wideband} at the point $(m-1)\tau=0$, which yields the linearized phase, given by \begin{equation}	\label{approximatephase}
\operatorname*{arg}([\bm{a}(\vartheta,\omega(n),n)]_m)\approx \varphi(n)-\omega(n)(m-1)\tau,
\end{equation}
where $\varphi(n)=\sum_{k=1}^{K}\phi_{k}(\Delta n)^k $, $\omega(n)=\sum_{k=1}^{K}k\phi_{k}(\Delta n)^{k-1} $ and $\tau=\frac{d}{c}\sin\vartheta$.
The approximation \eqref{approximatephase} is tight, if the Taylor approximation error is small, i.e., when the inequality
${\omega'(n)(m-1)^2\tau^2}/2 \ll \omega(n)(m-1)\tau$ with $\omega'(n)=\sum_{k=1}^K k(k-1)\phi_{k}(\Delta n)^{k-2}$ is satisfied \cite{rudin1976}. Therefore, coherent processing with PPSs can be achieved within the space-time interval $M_{\rm c, 1} \leq m \leq M_{\rm c,2}$ and $N_{\rm c,1}\leq n \leq N_{\rm c,2}$ in the case\begin{equation}
	\frac{\overline{\omega'}d(M_{\rm c,2}-1)}{ 2\underline{\omega}c} \ll 1,
\label{approximation}
\end{equation}
where  $\underline{\omega}=\min\limits_{n}\omega(n)$, $\overline{\omega'}=\max\limits_n|\omega'(n)|$ for $N_{\rm c,1}\leq n \leq N_{\rm c,2}$ and $\tau=\frac{d}{c}$ represent the worst case error bounds.

According to \eqref{beampatterna}, the largest magnitude of the beamformer output over the field of view, i.e., ${|P_b(\vartheta,\omega)|=1}$, is obtained, if $\left(\frac{\omega}{2\pi}\frac{d}{c}\sin\vartheta+\frac{b}{M}\right)\in\mathbb{Z}$.
Hence, the angular location of the mainlobe of the $b$th beamformer for $b=0,\ldots,M-1$ in dependence of the instantaneous frequency $\omega$ is given by the expression
\begin{equation}\label{DOAvsbeam}
	\vartheta_{\text{MB}}(\omega,b)=\begin{cases}
		\arcsin\left(-\frac{2\omega_0}{\omega}\frac{b}{M}\right), &\quad\text{for }b=0,\ldots,\frac{M}{2}-1,\\
		\arcsin\left(\frac{2\omega_0}{\omega}(1-\frac{b}{M})\right), &\quad\text{for } b={\frac{M}{2}},\ldots,{M-1},
	\end{cases}
\end{equation}
where the cut-off instantaneous frequency $\omega_0=\frac{\pi c}{d}$ corresponds to $d=\frac{\lambda}{2}$ with $\lambda$ denoting the wavelength of the signal \cite{vantrees2002}.

\begin{figure}[t]
\centering
\subcaptionbox{Beamformer magnitude response $|P_b(\vartheta,\omega)|$ (solid) and angular positions of maximum overlap of adjacent mainlobes $\vartheta_\text{MB}(\omega,b\pm\frac{1}{2})$ (dashed) for normalized instantaneous frequency $\frac{\omega}{\omega_0}=0.9$. \label{freqbeamformera}}
[\linewidth]{
\begin{tikzpicture}
\begin{axis}[
xtick={-90,-45,0,45,90},
xlabel={$\vartheta$ [deg]},
ylabel={$|P_b(\vartheta,\omega)|$},
grid=major,
ytick={0,1},
height=4cm,
width=0.45\textwidth,
xlabel near ticks,
ylabel near ticks,
ylabel shift={-0.1 cm},
enlargelimits=false,
legend style={font=\small,anchor=south,legend columns=6,at={(0.5,1.2)},draw=none},
]
\addlegendentry{$b=0$}\addlegendentry{$b=1$}\addlegendentry{$b=2$}\addlegendentry{$b=3$}

\addplot[ultra thick, blue,mark size=1pt] table [ col sep=space,x index=0,y index=1] {beampattern0.csv};
\addplot[ultra thick, orange,mark size=1pt] table [ col sep=space,x index=0,y index=1] {beampattern1.csv};
\addplot[ultra thick, green,mark size=1pt] table [ col sep=space,x index=0,y index=1] {beampattern2.csv};
\addplot[ultra thick, red,mark size=1pt] table [ col sep=space,x index=0,y index=1] {beampattern3.csv};
\addplot[thick,blue]  coordinates {(-16,0)(-16,0.4188854381999832)};
\addplot[thick, blue]  coordinates {(16,0)(16,0.4188854381999832)};
\addplot[thick, dashed,orange]  coordinates {(-16,0)(-16,0.4188854381999832)};
\addplot[thick, orange]  coordinates {(-56,0)(-56,0.4188854381999832)};
\addplot[thick,dashed, green]  coordinates {(16,0)(16,0.4188854381999832)};
\addplot[thick, green]  coordinates {(56,0)(56,0.4188854381999832)};
\addplot[thick,dashed,red]  coordinates {(-56,0)(-56,0.4188854381999832)};
\addplot[thick,dashed,red]  coordinates {(56,0)(56,0.4188854381999832)};
\end{axis}
\end{tikzpicture}
}
\subcaptionbox{Angular positions of beamformer mainlobes $\vartheta_\text{MB}(\omega,b)$ (solid) and angular positions of maximum overlap of adjacent beamformer mainlobes $\vartheta_\text{MB}(\omega,b\pm\frac{1}{2})$ (dashed) for normalized instantaneous frequencies $\frac{\omega}{\omega_0}\in[0,2]$. \label{freqbeamformerb}}
[\linewidth]{
\begin{tikzpicture}
\begin{axis}[
width=0.45\textwidth,
height=6cm,xmin=0, xmax=1000,
ymin=-90,ymax=90,
enlargelimits=false,
ylabel={$\vartheta$ [deg]},
ytick={-90,-45,0,45,90},
xlabel={$\frac{\omega}{\omega_0}$},
xlabel style={yshift=0.2cm},
xtick={0,250 ,500,750, 1000},
xticklabels={0, $\frac{1}{2}$,$1$, $\frac{3}{2}$,$2$},
mark=none,
grid=major,
ylabel near ticks,
xlabel near ticks,
mark repeat=2,
xlabel style={yshift=0.2cm},
ylabel style={yshift=-0.2cm},
legend style={font=\small,anchor=west,at={(0,0.42)}},
cycle list={
{ultra thick,mark size=1pt},
{thick, loosely dotted},
{thick, loosely dashed},
{ultra thick,mark size=1pt},
{thick, loosely dotted},
{thick, loosely dashed},
},
]

\addplot[ultra thick, blue,mark size=1pt] table [ col sep=space,x index=0,y index=1] {beam0.csv};
\addplot[thick, loosely dotted,blue] table [ col sep=space,x index=0,y index=2] {beam0.csv};
\addplot[thick, loosely dashed,blue] table [ col sep=space,x index=0,y index=3] {beam0.csv};
\addplot[ultra thick,blue,mark size=1pt] table [ col sep=space,x index=0,y index=4] {beam0.csv};
\addplot[thick, loosely dotted,blue] table [ col sep=space,x index=0,y index=5] {beam0.csv};
\addplot[thick, loosely dashed,blue] table [ col sep=space,x index=0,y index=6] {beam0.csv};

\addplot[ultra thick, orange,mark size=1pt] table [ col sep=space,x index=0,y index=1] {beam1.csv};
\addplot[thick, loosely dotted,orange] table [ col sep=space,x index=0,y index=2] {beam1.csv};
\addplot[thick, loosely dashed,orange] table [ col sep=space,x index=0,y index=3] {beam1.csv};
\addplot[ultra thick,orange,mark size=1pt] table [ col sep=space,x index=0,y index=4] {beam1.csv};
\addplot[thick, loosely dotted,orange] table [ col sep=space,x index=0,y index=5] {beam1.csv};
\addplot[thick, loosely dashed,orange] table [ col sep=space,x index=0,y index=6] {beam1.csv};

\addplot[ultra thick, green,mark size=1pt] table [ col sep=space,x index=0,y index=1] {beam2.csv};
\addplot[thick, loosely dotted,green] table [ col sep=space,x index=0,y index=2] {beam2.csv};
\addplot[thick, loosely dashed,green] table [ col sep=space,x index=0,y index=3] {beam2.csv};
\addplot[ultra thick,green] table [ col sep=space,x index=0,y index=4] {beam2.csv};
\addplot[thick, loosely dotted,green] table [ col sep=space,x index=0,y index=5] {beam2.csv};
\addplot[thick, loosely dashed,green] table [ col sep=space,x index=0,y index=6] {beam2.csv};

\addplot[ultra thick, red,mark size=1pt] table [ col sep=space,x index=0,y index=1] {beam3.csv};
\addplot[thick, loosely dotted,red] table [ col sep=space,x index=0,y index=2] {beam3.csv};
\addplot[thick, loosely dashed,red] table [ col sep=space,x index=0,y index=3] {beam3.csv};
\addplot[ultra thick,red,mark size=1pt] table [ col sep=space,x index=0,y index=4] {beam3.csv};
\addplot[thick, loosely dotted,red] table [ col sep=space,x index=0,y index=5] {beam3.csv};
\addplot[thick, loosely dashed,red] table [ col sep=space,x index=0,y index=6] {beam3.csv};
\end{axis}
\end{tikzpicture}
}
\caption{Beamformer magnitude response $|P_b(\vartheta,\omega)|$ over the field-of-view $\vartheta\in[-\frac{\pi}{2},\frac{\pi}{2}]$ for $M=4$ DFT beamformers, beam index $b=0,\ldots,M-1$ and normalized instantaneous frequencies $\frac{\omega}{\omega_0}\in[0,2]$ with cut-off instantaneous frequency $\omega_0=\frac{\pi c}{d}$ \label{freqbeamformer}}
\end{figure}
 Fig.~\ref{freqbeamformer} shows the magnitude response $|P_b(\omega,\vartheta)|$ across the field of view $\vartheta\in[-\frac{\pi}{2},\frac{\pi}{2}]$ for $M=4$ DFT beamformers, illustrating the conditions for coherent array processing with PPS.
In Fig.~\ref{freqbeamformera}, the magnitude response is displayed over the field of view for the normalized instantaneous frequency $\frac{\omega}{\omega_0}=0.9$.
The angles $\vartheta_{\text{MB}}(\omega,b\pm\frac{1}{2})$ indicate the locations of maximum overlap between adjacent beamformer mainlobes. 
Fig.~\ref{freqbeamformerb} illustrates how the angular positions of the beamformer mainlobes $\vartheta_{\text{MB}}(\omega,b)$ and the maximum overlap between adjacent mainlobes $\vartheta_{\text{MB}}(\omega,b\pm\frac{1}{2})$ vary with respect to the normalized instantaneous frequencies $\frac{\omega}{\omega_0}\in[0,2]$.
The angular location of the beamformer mainlobe $\vartheta_{\text{MB}}(\omega,0)$ is constant at $\vartheta=\cmddegree{0}$ for all instantaneous frequencies, whereas, the angular locations of the other mainlobes, $\vartheta_{\text{MB}}(\omega,b)$ for $b=1,\ldots,M-1$, vary with changing instantaneous frequency.
When $\frac{\omega}{\omega_0} > 1$, spatial aliasing may occur due to grating lobes in the beamformer output, potentially causing ambiguities in the DOA estimation \cite{vantrees2002}.
Moreover, a limitation of time-invariant beamforming with PPSs is that as the instantaneous frequency of the impinging signal changes, the signal energy can leak into adjacent DFT beamformers.
Therefore, a change in the instantaneous frequency of the impinging signal is associated with a change in its spatial signature in the beamformer output.

\subsection{Sparse Support of Time-Frequency-Spatial Signatures}\label{sparsesupport}
In the following, we define the sparse support of the time-frequency spatial source signatures. 
The RANSAC method is applied to this sparse support set to estimate the source parameters.

For each beamformer $b$, the set of local maxima in both frequency and beam dimensions is determined as
\begin{equation}
	\mathcal{S}_b=\Big\{z_b(p,q):\left|z_b (p,q)\right|\ge\max\limits_{\substack{q'\in\{q-1,q+1\}\\b'\in\{b-1,b+1\}}} \left|z_{b'}(p,q')\right|\Big\}\label{S_b}.
\end{equation}
Furthermore, a thresholding operation is applied to the set $\mathcal{S}_b$ as
\begin{equation}
	\mathcal{T}_b=\left\{(p,q)\in\mathcal{S}_b : |z_b(p,q)|\ge \epsilon\right\}\label{T_b},
\end{equation}
to select the indices of the time-frequency points in the set $\mathcal{S}_b$ with a magnitude that is equal or larger to the threshold $\epsilon\in\Rbb$.

\begin{figure}[t]
\centering
\begin{tikzpicture}
\begin{groupplot}[
group style ={group size =4 by 4,
vertical sep=0cm,horizontal sep=0cm,},
width=3.2cm,height=3.2cm,
xmin=-0.56,xmax=0.56,
ymin=-314,ymax=314,
xtick={-0.35,0.35},
ytick={-220,220},
yticklabels={-0.3,0.3},
xticklabels={-0.5,0.5},
ymajorticks=false,
xmajorticks=false,
ylabel near ticks,
xlabel near ticks,
xlabel shift={-0.1cm},
ylabel shift={-0.2 cm},
title style={yshift=-0.2cm},
ytick style={draw=none},xtick style={draw=none},
]
\nextgroupplot[title={$b=0$},
ymajorticks=true,
ylabel style={align=center},
ylabel={$|y_m(p,q)|$\\$\omega$ [kHz]},]
\addplot graphics [xmin=-0.56,xmax=0.56,ymin=-314,ymax=314] {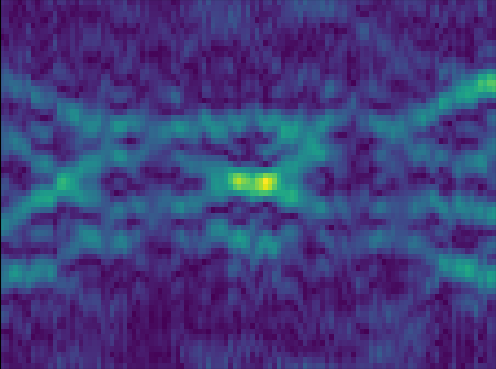} ;
\nextgroupplot[title={$b=1$},]\addplot graphics [xmin=-0.56,xmax=0.56,ymin=-314,ymax=314] {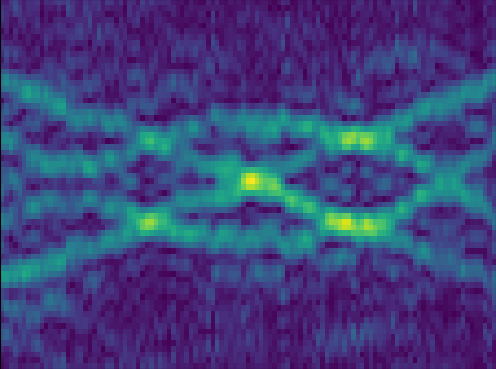};
\nextgroupplot[title={$b=2$},]\addplot graphics [xmin=-0.56,xmax=0.56,ymin=-314,ymax=314] {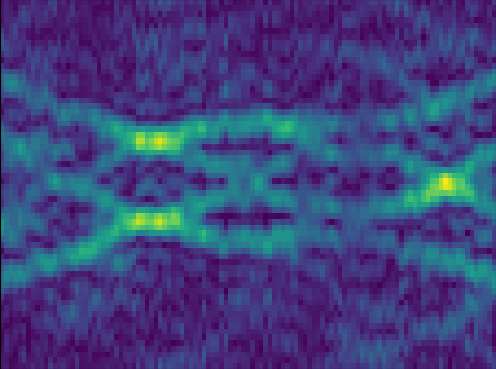};
\nextgroupplot[title={$b=3$},ylabel={a$)$},ylabel style={yshift=-1.9cm,rotate=-90}
]\addplot graphics [xmin=-0.56,xmax=0.56,ymin=-314,ymax=314]{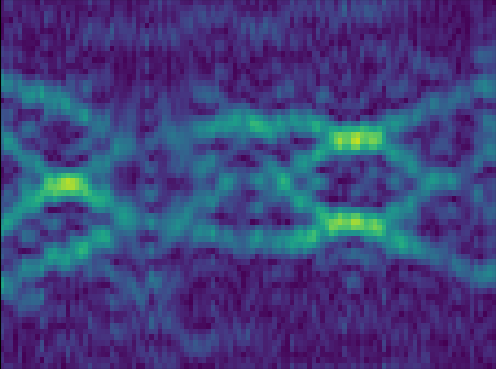};
\nextgroupplot[ymajorticks=true,ylabel style={align=center},ylabel={$|z_b(p,q)|$\\$\omega$ [kHz]},]
\addplot graphics [xmin=-0.56,xmax=0.56,ymin=-314,ymax=314] {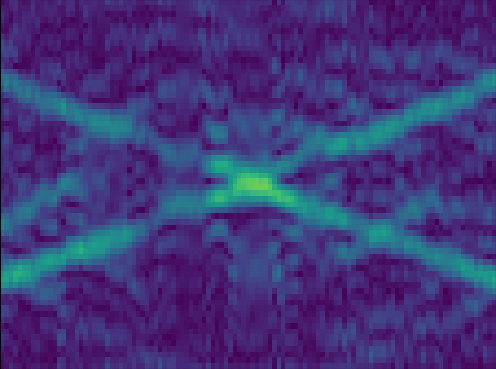} ;
\nextgroupplot[]\addplot graphics [xmin=-0.56,xmax=0.56,ymin=-314,ymax=314] {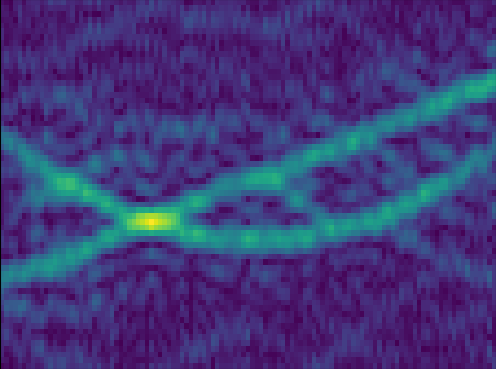};
\nextgroupplot[]\addplot graphics [xmin=-0.56,xmax=0.56,ymin=-314,ymax=314] {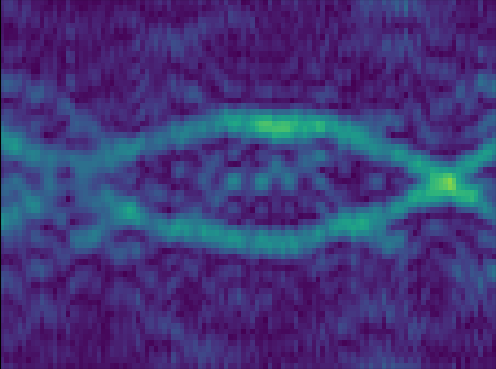};
\nextgroupplot[	ylabel={b$)$},ylabel style={yshift=-1.9cm,rotate=-90}]
\addplot graphics [xmin=-0.56,xmax=0.56,ymin=-314,ymax=314]{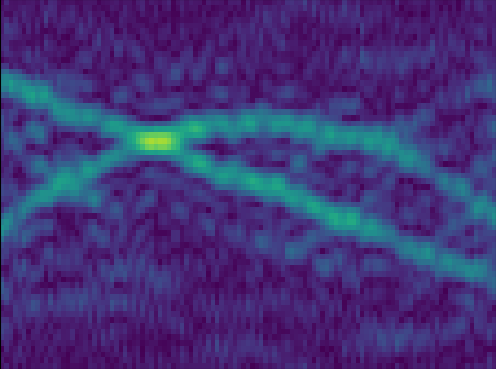};
\nextgroupplot[ymajorticks=true,ylabel style={align=center},ylabel={$\mathcal{S}_b$\\$\omega$ [kHz]},]
\addplot [scatter,mark size=1pt,only marks,]  table [col sep=space,x index=0,y index=1,scatter src=\thisrowno{2}] {S0.csv};
\nextgroupplot[]\addplot [scatter,mark size=1pt,only marks,]  table [col sep=space,x index=0,y index=1,scatter src=\thisrowno{2}] {S1.csv};
\nextgroupplot[]\addplot [scatter,mark size=1pt,only marks,]  table [col sep=space,x index=0,y index=1,scatter src=\thisrowno{2}] {S2.csv};
\nextgroupplot[	ylabel={c$)$},ylabel style={yshift=-1.9cm,rotate=-90}]
\addplot [scatter,mark size=1pt,only marks,]  table [col sep=space,x index=0,y index=1,scatter src=\thisrowno{2}]{S3.csv};
\nextgroupplot[	xlabel={$t$ [s]},xmajorticks=true,ymajorticks=true,ylabel style={align=center},ylabel={$\mathcal{T}_b$\\$\omega$ [kHz]},]
\addplot [mark size=1pt,only marks,black]table [col sep=space,x index=0,y index=1] {T0.csv};
\nextgroupplot[,xmajorticks=true,xlabel={$t$ [s]},]
\addplot [mark size=1pt,only marks,black]table [col sep=space,x index=0,y index=1] {T1.csv};
\nextgroupplot[xmajorticks=true,xlabel={$t$ [s]},]
\addplot [mark size=1pt,only marks,black]table [col sep=space,x index=0,y index=1] {T2.csv};
\nextgroupplot[xmajorticks=true,xlabel={$t$ [s]},ylabel={d$)$},ylabel style={yshift=-1.9cm,rotate=-90}]
\addplot [mark size=1pt,only marks,black]table [col sep=space,x index=0,y index=1]{T3.csv};
\end{groupplot}
\end{tikzpicture}
\caption{Data processing steps to obtain the sparse support of the time-frequency spatial signatures of the four PPS sources specified in Table~\ref{table} and an array consisting of four sensors.
a) Time-frequency spectra according to STFT in \eqref{y_m} at $\text{SNR}=0~\text{dB}$,
b) Time-frequency spectra at DFT beamformer output in \eqref{z_b},
c) Time-frequency points in set $\mathcal{S}_b$ obtained after local maximum search in \eqref{S_b},
d) Time-frequency points in set $\mathcal{T}_b$ obtained after thresholding step in \eqref{T_b}. \label{preprocessing}}
\end{figure}
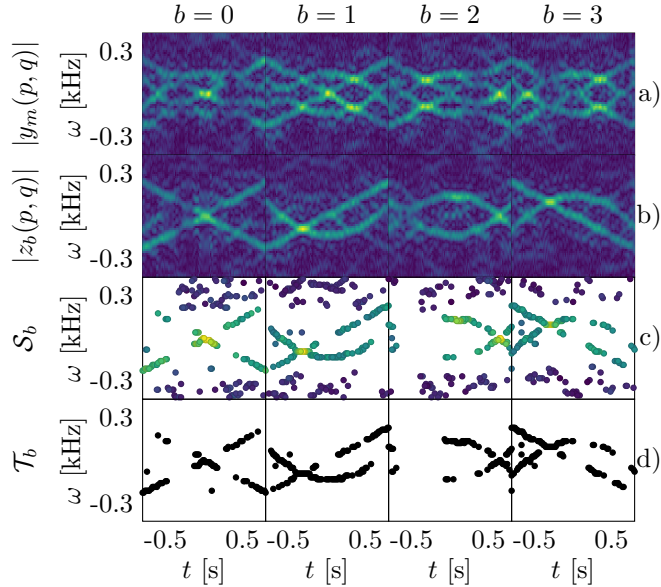
 The data processing steps described in Sections \ref{sec:dftbeamformers} and \ref{sparsesupport} are illustrated in Fig.~\ref{preprocessing} for an example of four PPSs specified in Table~\ref{table} impinging on an array consisting of $M=4$ sensors.
Fig.~\ref{preprocessing}a shows the STFT-based time-frequency spectra of \eqref{y_m} across different sensors, revealing similar patterns in each sensor.
Fig.~\ref{preprocessing}b presents the time-frequency spectra at the DFT beamformer outputs in \eqref{z_b}, where different beamformers capture the PPSs according to their spatial signatures.
Signal components may appear in time-frequency spectra of adjacent beamformers due to energy leakage, referring to the aforementioned limitation of time-invariant beamforming when processing signals with time-varying instantaneous frequencies.
Fig.~\ref{preprocessing}c displays the time-frequency points in the set $\mathcal{S}_b$ for $b=0,\ldots,M-1$. The time-frequency points of the signals are preserved and the redundant time-frequency signatures in adjacent beamformers are effectively removed by the local maximum search.
However, local maxima of low magnitude remain in $\mathcal{S}_b$, which correspond to noise components as well as spectral leakage of the STFT and spatial leakage of the beamformer sidelobes.
Fig.~\ref{preprocessing}d displays the time-frequency points at the indices in the set $\mathcal{T}_b$ for $b=0,\ldots,M-1$, where only the data points of the true signal components are preserved.
Hence, the sparse support set of the time-frequency spatial source signatures is given by the union $\mathcal{T}_0\cup\cdots\cup\mathcal{T}_{M-1}$.

For a single PPS of polynomial order $K$, the polynomial coefficients can be uniquely determined from $K$ time-frequency points that lie on the time-frequency signature of the signal \cite{djurovic2019}.
Hence, the RANSAC method can be used for PPS parameter estimation by randomly sub-sampling $K$ local maxima in the time-frequency spectrum  and matching a $K$th order polynomial to their corresponding time-frequency locations \cite{djurovic2017}.
In the multi-source case, the time-frequency signature of $L$ signals overlap, hence, $L$ polynomials with distinct polynomial coefficients need to be simultaneously matched to estimate the PPS parameters of $L$ sources.
However, the resulting estimation problem is combinatorial and the probability of sub-sampling $K$ points from the same signature for each source decreases with the number of signals and the polynomial order.
When the number of sources $L$ is large, the number of required sub-samples to obtain appropriate parameter estimates for all sources becomes unacceptable. 

 Therefore, we propose estimating the DOA and PPS parameters of one source at a time in a sequential estimation procedure using the RANSAC method  and employ the $i$-source ML approximation criterion in \eqref{isource} to select the best-fitting candidate parameter estimates.
According to Fig.~\ref{preprocessing}d, the sparse support of a particular signal may be distributed across adjacent beamformers.
Therefore, in the RANSAC procedure of the $i$th sequential iteration, we consider the union, consisting of the set $\mathcal{T}_{b^{(i)}}$ and its adjacent sets, given by
\begin{equation}
	\mathcal{T}^{(i)}= \{\mathcal{T}_{b^{(i)}-1}\cup\mathcal{T}_{b^{(i)}} \cup \mathcal{T}_{b^{(i)}+1}\}.\label{union}
\end{equation}
The beam index $b^{(i)}$ refers to the beamformer, which captures the highest signal energy among all beamformers, hence, the particular set $\mathcal{T}_b$ in \eqref{T_b} for $b=0,\ldots,M-1$ with the largest cardinality, given by
\begin{equation}
	{b}^{(i)}=\argmax\limits_{b\in\{0,\ldots,M-1\}}  |\mathcal{T}_b|\label{RANSAC0}.
\end{equation}

\subsection{Source Parameter Estimation\label{RANSAC}}

In the following, the iterative RANSAC method for estimating the DOA and PPS parameters of one source at a time in the proposed sequential estimation procedure is explained \cite{fischler1981,djurovic2017}.

 \subsubsection{RANSAC}
 In each RANSAC iteration, $K$ time-frequency points are sub-sampled at random from the set $\mathcal{T}\cmdi$ in \eqref{union} and the corresponding PPS parameters are estimated by solving a system of $K$ equations with $K$ unknowns given by
\begin{equation}\label{polynomialregression}
	\begin{bmatrix}
		\omega(q\cmdl_1)\\
\vdots		\\
		\omega(q\cmdl_K)
	\end{bmatrix}
	=
	\begin{bmatrix}
		1 & 2t(p\cmdl_1)&\ldots&	Kt(p\cmdl_1)^{K-1}\\
\vdots & \vdots&\ddots&\vdots\\
		1 & 2t(p\cmdl_K)&\ldots&	K t(p\cmdl_K)^{K-1}\\
	\end{bmatrix}
	\begin{bmatrix}
		\phi\cmdl_{1}\\
\vdots		\\
		\phi\cmdl_{K}
	\end{bmatrix},
\end{equation}
where the corresponding least-squares (LS) solution yields the PPS parameter estimate $\hat{\cmdp}\cmdl=\left[\hat\phi\cmdl_{1},\ldots,\hat\phi\cmdl_{K}\right]\trans$.
The superscript $(\cdot)\cmdl$ denotes the $\lambda$th RANSAC iteration in the $i$th sequential iteration for $\lambda=1,\ldots,\Lambda$, where $\Lambda\in\mathbb{N}$ is the number of RANSAC iterations.
The time and frequency values at the time index $p=-\left\lceil\frac{N-H}{2}\right\rceil,\ldots,\left\lfloor\frac{N-H}{2}\right\rfloor$ and frequency index $q=-\left\lceil\frac{H-1}{2}\right\rceil,\ldots,\left\lfloor\frac{H-1}{2}\right\rfloor$ of the STFT in \eqref{y_m}, are given by\begin{align}\label{timefrequency}
	\begin{split}
		t(p)&=\Delta p,\\
		\omega(q)&=\frac{2\pi q}{\Delta H}.
	\end{split}
\end{align}
To ensure good conditioning for the polynomial interpolation \eqref{polynomialregression}, the $K$ time instants $t(p\cmdl_1),\ldots,t(p\cmdl_K)$ are sampled from $K$ non-overlapping, equal partitions of the time axis.
For the $k$th partition for $k=1,\ldots,K$, this corresponds to the set
\begin{equation}\label{partition}
	{\mathcal{P}}_k=\{(p,q)\in\mathcal{T}^{(i)}| (k-1)\overline{N}\le p<k \overline{N}\},
\end{equation}
where $\overline{N}=\left\lceil\frac{N-H}{K}\right\rceil$. The time and frequency index $(p\cmdl_k,q\cmdl_k)$ in \eqref{polynomialregression} is thus sampled from ${\mathcal{P}}_k$, according to the discrete uniform distribution $\mathcal{U}\{1,|\mathcal{P}_k]\}$ for $k=1,\ldots,K$.

The source angle $\hat\theta\cmdl$, corresponding to the PPS parameter estimate $\hat\cmdp\cmdl$ from \eqref{polynomialregression}, is determined by performing a spectral search on the $i$-source ML approximation in \eqref{isource} as\begin{equation}
	\hat\theta\cmdl=\argmin\limits_{\vartheta\in{\mathcal{Q}\cmdl}}f_i(\vartheta,\hat{\cmdp}\cmdl)\label{RANSACtheta}.
\end{equation}
The angular interval for the spectral search in \eqref{RANSACtheta} is determined using the function $\vartheta_{\text{MB}}(\omega,b)$ in \eqref{DOAvsbeam}, given by\begin{equation}
	\mathcal{Q}\cmdl=\left[\min \limits_{n}\vartheta_{\text{MB}}(\hat{\omega}\cmdl(n),b^{(i)}),\max \limits_{n}\vartheta_{\text{MB}}(\hat{\omega}\cmdl(n),b^{(i)})\right],\label{Q}
\end{equation}
where $b\cmdi$ and $\hat{\omega}\cmdl(n)=\sum_{k=1}^{K}k\hat{\phi}\cmdl_{k} (\Delta n)^{k-1}$ are the beam index in \eqref{RANSAC0} and the instantaneous frequency corresponding to the PPS parameter estimate from \eqref{polynomialregression}, respectively.

To reduce the computational complexity in \eqref{RANSACtheta}, the efficient implementation of the $i$-source ML approximation in \eqref{isource} is used, in which, the estimated source parameters in the previous $(i-1)$ sequential iterations are assumed to be fixed.
As shown in \ref{appendix:isource} \cite{trinh-hoang2018}, this efficient $i$-source ML approximation is given by
\begin{equation}\label{isource:simplified}
	f_i(\vartheta,\hat{\cmdp}\cmdl)=\|\cmdr\cmdim\|_2^2-\frac{|\bar{\bm{a}}^\mathsf{H}(\vartheta,\hat{\cmdp}\cmdl)\cmdr\cmdim|}{\|\hat{\bm{a}}(\vartheta,\hat{\cmdp}\cmdl)\|_2^2},
\end{equation}
where $\hat{\bm{a}}(\vartheta,\hat{\cmdp}\cmdl)=\bm{\Pi}_{\bar {\bm A}}^\perp(\hat{\bm{\theta}}\cmdim,\hat{\cmdP}\cmdim)\bar{\bm{a}}(\vartheta,\hat{\cmdp}\cmdl)$. It is worth noting that the orthogonal projection matrix $\bm{\Pi}_{\bar {\bm A}}^\perp(\hat{\bm{\theta}}\cmdim,\hat{\cmdP}\cmdim)$ is already available from the computation of the residual received signal $\cmdr\cmdim$ in \eqref{residual}.

\subsubsection{Parameter Refinement}
After $\Lambda$ RANSAC iterations, the DOA and PPS parameter estimates of the $i$th source are determined as 
\begin{equation}
	\{\hat\theta^{(i)},\hat{\cmdp}^{(i)}\}=\{\hat\theta^{(i,\lambda^\star)},\hat{\cmdp}^{(i,\lambda^\star)}\}\label{RANSAC2},
\end{equation}
which correspond to the best-fitting candidate parameter estimates, found in the $(\lambda^\star)$th RANSAC iteration, such that\begin{equation}
	\lambda^\star=\argmin_{\lambda\in\{1,\ldots,\Lambda\}}f_i(\hat\theta\cmdl,\hat{\cmdp}\cmdl).\label{lambdamax}
\end{equation}

It is worth noting that the source parameter estimates of the $i$th source in \eqref{RANSAC2} are only coarse estimates of the true source parameters.
This is because the accuracy of $\hat{\cmdp}^{(i)}$ is limited by the frequency resolution $\frac{2\pi}{\Delta H}$ of the STFT in \eqref{y_m} and the accuracy of $\hat\theta^{(i)}$ is limited by the angular grid resolution of $\mathcal{Q}\cmdl$ in \eqref{Q}.
Therefore, a gradient-based parameter refinement step is performed subsequently to the RANSAC method.
This refinement is a local gradient search on the original $i$-source ML approximation in \eqref{isource} as
\begin{equation}
\{\hat\cmdt^{(i)},\hat\cmdP^{(i)}\}=\argmin\limits_{\cmdt,\cmdP} f_i(\cmdt,\cmdP)\label{BFGS},
\end{equation}
where the initial values $\cmdt_0=\left[\hat{\cmdt}\cmdim,\hat\theta^{(i)}\right]\in\Rbb^{i}$ and $\cmdP_0=\left[\hat{\cmdP}\cmdim,\hat\cmdp^{(i)}\right]\in\Rbb^{K\times i}$ are obtained from the source parameter estimates in the previous $(i-1)$ sequential iterations and the source parameter estimates of the $i$th source in \eqref{RANSAC2}.
The gradient of the original $i$-source ML approximation $f_i(\cmdt,\cmdP)$ in \eqref{isource} is provided in \ref{appendix:gradient}. 

Subsequently, the residual received signal $\cmdr^{(i)}$ is obtained, according to \eqref{residual}, using the refined parameter estimates $\{\hat\cmdt^{(i)},\hat\cmdP^{(i)}\}$ of the $i$ sources in \eqref{BFGS}.
In the $(i+1)$th sequential iteration, the residual received signal $\cmdr^{(i)}$ is then processed in \eqref{y_m} to effectively form the sparse support $\mathcal{T}^{(i+1)}$ in \eqref{union} of the remaining sources.
The sequential DOA-PPS estimator described in Sections \ref{sec:dftbeamformers} - \ref{RANSAC} is summarized in Alg.~\ref{killer}.
\begin{algorithm}[t]
\caption{Sequential DOA-PPS Estimator\label{killer}}
\begin{algorithmic}[1]
\State {\textit{Initialization} $\cmdr^{(0)}=\bar{\bm{x}}$ $\gets$ Eq.~\eqref{stacked}}
\For{\textit{Source iterations} $i=1,\ldots,L$}
\State \textit{Compute time-frequency beamspace representation $\mathcal{T}_0,\ldots,\mathcal{T}_{M-1}$ of residual $\cmdr\cmdim$} $\gets$ Eq.~\eqref{T_b}\State \textit{Compute sparse support set for dominant beam} $\mathcal{T}^{(i)}$ $\gets$ Eq.~\eqref{union}\For{\textit{RANSAC iterations} $\lambda=1,\ldots,\Lambda$ }
\State \textit{Estimate candidate PPS parameters} $\hat{\cmdp}\cmdl$ $\gets$ Eq.~\eqref{polynomialregression}\State \textit{Estimate candidate DOA} $\hat\theta\cmdl$ $\gets$ Eq.~\eqref{RANSACtheta}\State \textit{Cost function evaluation using $i$-source ML approximation} $f_i(\hat{\theta}\cmdl,\hat{\cmdp}\cmdl)$ $\gets$ Eq.~\eqref{isource:simplified}
\EndFor
\State \textit{Select parameter estimates that minimize $i$-source ML approximation} $\hat{\theta}^{(i)},\hat{\cmdp}^{(i)}$ $\gets$ Eq.~\eqref{RANSAC2}, \eqref{lambdamax}
\State \textit{Parameter refinement} $\hat\cmdt^{(i)},\hat{\cmdP}^{(i)}$ \textit{based on $i$-source ML approximation} $f_i(\cmdt,\cmdP)$ $\gets$ Eq.~\eqref{BFGS} 
\State \textit{Signal elimination in raw data via residual} $\cmdr^{(i)}$ $\gets$ Eq.~$\eqref{residual}$
\EndFor\\
\Return  $\hat\cmdt^{(L)},\hat{\cmdP}^{(L)}$
\end{algorithmic}
\end{algorithm}
\subsection{Low-Cost Sequential DOA-PPS Estimator\label{lowcost}}
We introduce a simplified variant of the sequential DOA-PPS estimator in Alg.~\ref{killer} that trades performance for reduced complexity.
The effects of each simplification on the estimation performance are investigated separately.
\subsubsection{Complexity Reduction for the Data Processing\label{lowcost:refinement}}
The processing of the measurement data in Alg.~\ref{killer}, described in Sections \ref{sec:dftbeamformers} - \ref{sparsesupport}, consists of obtaining the sparse support $\mathcal{T}\cmdi$ in \eqref{union} by processing the residual received signal $\bm{r}\cmdim$ in \eqref{residual} in each sequential iteration.
The total complexity of computing the STFTs in \eqref{y_m} is thus of order $\mathcal{O}(ML(N-H)H\log H)$.

\begin{figure}[t]
\centering
\begin{tikzpicture}
\begin{groupplot}[
group style ={group size =4 by 1,
vertical sep=0cm,horizontal sep=0cm,},
width=3.2cm,
height=3.2cm,
xmin=-0.56,xmax=0.56,
ymin=-314,ymax=314,
xtick={-0.35,0.35},
ytick={-220,220},
yticklabels={-0.3,0.3},
xticklabels={-0.5,0.5},
ymajorticks=false,
xmajorticks=false,
ylabel near ticks,
xlabel near ticks,
xlabel shift={-0.1cm},
ylabel shift={-0.2 cm},
title style={yshift=-0.2cm},
ytick style={draw=none},xtick style={draw=none},
]
\nextgroupplot[	xlabel={$t$ [s]},xmajorticks=true,ymajorticks=true,ylabel style={align=center},ylabel={$\omega$ [kHz]},]
\addplot [mark size=1pt,only marks,black]table [col sep=space,x index=0,y index=1] {T0.csv};
\addplot[red,ultra thick,dashed]table[col sep=space,x index=0,y index=8]{corridor.csv};
\addplot[red,ultra thick,dashed]table[col sep=space,x index=0,y index=12]{corridor.csv};
\nextgroupplot[,xmajorticks=true,xlabel={$t$ [s]},]
\addplot [mark size=1pt,only marks,black]table [col sep=space,x index=0,y index=1] {T1.csv};
\addplot[red,ultra thick,dashed]table[col sep=space,x index=0,y index=5]{corridor.csv};
\addplot[red,ultra thick,dashed]table[col sep=space,x index=0,y index=9]{corridor.csv};
\nextgroupplot[xmajorticks=true,xlabel={$t$ [s]},]
\addplot [mark size=1pt,only marks,black]table [col sep=space,x index=0,y index=1] {T2.csv};
\addplot[red,ultra thick,dashed]table[col sep=space,x index=0,y index=7]{corridor.csv};
\addplot[red,ultra thick,dashed]table[col sep=space,x index=0,y index=11]{corridor.csv};
\nextgroupplot[xmajorticks=true,xlabel={$t$ [s]}]
\addplot [mark size=1pt,only marks,black]table [col sep=space,x index=0,y index=1]{T3.csv};
\addplot[red,ultra thick,dashed]table[col sep=space,x index=0,y index=6]{corridor.csv};
\addplot[red,ultra thick,dashed]table[col sep=space,x index=0,y index=10]{corridor.csv};
\end{groupplot}
\end{tikzpicture}
\caption{Sparse support of the time-frequency spatial signatures and the margins (red) $\left\{\hat{\omega}^{(i)}(n)+\frac{2\pi}{\Delta H},~\hat{\omega}^{(i)}(n)-\frac{2\pi}{\Delta H}\right\}$ corresponding to the set $\mathcal{C}\cmdi$ in \eqref{corridor0} for $i=1,\ldots,L$.\label{fig:corridor}}
\end{figure}
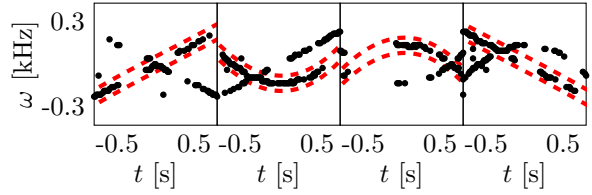

In this simplified variant, source elimination is performed directly in the time-frequency-beamspace domain rather than on the original measurements.
Consequently, the processing steps in Sections \ref{sec:dftbeamformers} - \ref{sparsesupport} are executed only once, reducing the STFT computation complexity to $\mathcal{O}(M(N-H)H\log H)$.
This is accomplished by determining the time-frequency points of the sparse support set $\mathcal{T}\cmdi$, which lie within one STFT frequency bin $\frac{2\pi}{\Delta H}$ around the estimated instantaneous frequency $\hat{\omega}^{(i)}(n)$ of the $i$th source \cite{djurovic2018a}.
In the $i$th sequential iteration, this corresponds to the set
\begin{equation}
		\mathcal{C}\cmdi=\Big\{(p_j,q_j)\in\mathcal{T}\cmdi:|\hat{\omega}^{(i)}(n)-\omega(q_j)|\le \frac{2\pi}{\Delta H},\Delta n=t(p_j)\Big\}\label{corridor0},
\end{equation}
where
${\hat{\omega}^{(i)}(n)=\sum_{k=1}^{K}k\hat{\phi}^{(i,\lambda^\star)}_{k} (\Delta n)^{k-1}}$
is formed from the PPS parameter estimates $\{\hat{\phi}^{(i,\lambda^\star)}_{1},\ldots,\hat{\phi}^{(i,\lambda^\star)}_{K}\}$ of the $i$th source in \eqref{RANSAC2}.
Afterwards, the $i$th estimated source is eliminated from $\mathcal{T}\cmdi$ by performing the set difference as
 \begin{equation}
 	\mathcal{T}^{(i+1)}=\mathcal{T}^{(i)}\setminus\mathcal{C}\cmdi. \label{removecorridor}
 \end{equation}
 In Fig.~\ref{fig:corridor}, the margins $\left\{\hat{\omega}^{(i)}(n)+\frac{2\pi}{\Delta H},\hat{\omega}^{(i)}(n)-\frac{2\pi}{\Delta H}\right\}$ and the set $\mathcal{C}\cmdi$ for $i=1,\ldots,L$ of the sparse support in Fig.~\ref{preprocessing}d are shown.

 To simplify the parameter refinement of Alg.~\ref{killer}, the local gradient search on the $i$-source ML approximation in \eqref{BFGS} is replaced by the LS refitting using the time-frequency points at the elements in the set $\mathcal{C}\cmdi$. This corresponds to the overdetermined system of $|\mathcal{C}\cmdi|>K$ equations with $K$ unknowns given by \begin{equation}\label{refitting}
\begin{bmatrix}
\omega(q_1)\\
\vdots		\\
\omega(q_{|\mathcal{C}\cmdi|})
\end{bmatrix}
\approx
\begin{bmatrix}
1 & 2t(p_1)&\ldots&	Kt(p_1)^{K-1}\\
\vdots & \vdots&\ddots&\vdots\\
1 & 2t(p_{|\mathcal{C}\cmdi|})&\ldots&	Kt(p_{|\mathcal{C}\cmdi|})^{K-1}\\
\end{bmatrix}
\begin{bmatrix}
\phi_1\cmdi\\
\vdots		\\
\phi_{K}\cmdi
\end{bmatrix},
\end{equation}
where the LS solution yields the refined PPS parameter estimate of the $i$th source $\hat\cmdp\cmdi=\left[\hat{\phi}\cmdi_1,\ldots,\hat{\phi}\cmdi_K\right]^\mathsf{T}$.

\subsubsection{Complexity Reduction for RANSAC\label{lowcost:RANSAC}}
In the RANSAC method of Alg.~\ref{killer}, the spectral search on the $i$-source ML approximation in \eqref{isource:simplified} over the angular interval $\mathcal{Q}\cmdl$ is performed in each RANSAC iteration.
Hence, the total complexity of applying the RANSAC method for the estimation of the parameters of the $L$ sources in Alg.~\ref{killer} is of order $\mathcal{O}(G\Lambda  L(MN)^2)$, where $G\ge|\mathcal{Q}\cmdl|$ represents the maximum number of elements in the angular interval $\mathcal{Q}\cmdl$ in \eqref{Q}.

In the simplified variant, we reduce this complexity by replacing the exhaustive spectral search over $\mathcal{Q}\cmdl$ with random uniform sampling, such that
\begin{equation}
	\hat\theta\cmdl=\vartheta\in\mathcal{Q}\cmdl, \quad \vartheta \sim  \mathcal{U}\{1,|\mathcal{Q}\cmdl|\}\label{random}.
\end{equation}
Furthermore, the polynomial-phase beamformer $f_1(\theta,\cmdp)$ in \eqref{polynomialbeamformer} is used to evaluate the candidate DOA and PPS parameter estimates, $\hat{\theta}\cmdl$ and $\hat{\cmdp}\cmdl$ in \eqref{random} and \eqref{polynomialregression}, respectively, thus replacing the $i$-source ML approximation criterion in \eqref{lambdamax},  such that
\begin{equation}
	\lambda^\star=\argmax_{\lambda\in\{ 1,\ldots,\Lambda\}}f_1(\hat\theta\cmdl,\hat{\cmdp}\cmdl). \label{RANSACsimple}
\end{equation}
The total RANSAC complexity of the low-cost sequential estimator is thus reduced to $\mathcal{O}(\Lambda LMN)$.

Moreover, in the low-cost sequential estimator, the local gradient search in \eqref{BFGS} is performed only once on the full-source ML cost function $f_L(\cmdt,\cmdP)$ in \eqref{likelihood}, where the initial values are the DOA and PPS parameter estimates $\{\hat{\theta}^{(i,\lambda^\star)},\hat{\cmdp}^{(i,\lambda^\star)}\}$ of the $L$ sources, according to \eqref{RANSAC2} and \eqref{RANSACsimple} for $i=1,\ldots,L$. The low-cost sequential DOA-PPS estimator is summarized in Alg.~\ref{simple}.

\begin{algorithm}[t]
\caption{Low-Cost Sequential DOA-PPS Estimator\label{simple}}
\begin{algorithmic}[1]
\State \textit{Initialization} $\bar{\bm{x}}$ $\gets$ Eq.~\eqref{stacked} \State \textit{Compute time-frequency beamspace representation} $\mathcal{T}_0,\ldots,\mathcal{T}_{M-1}$ \textit{of} $\bar{\bm{x}}$ $\gets$ Eq.~\eqref{T_b}\For{\textit{Source iterations} $i=1,\ldots,L$}
\State \textit{Compute sparse support set for dominant beam} $\mathcal{T}^{(i)}$ $\gets$ Eq.~\eqref{union}\For{\textit{RANSAC iterations} $\lambda=1,\ldots,\Lambda$}
\State \textit{Estimate candidate PPS parameters} $\hat{\cmdp}\cmdl$ $\gets$ Eq.~\eqref{polynomialregression}\State \textit{Estimate candidate DOA} $\hat\theta\cmdl$ $\gets$ Eq.~\eqref{random}\State \textit{Cost function evaluation using single-source ML approximation} $f_1(\hat{\theta}\cmdl,\hat{\cmdp}\cmdl)$ $\gets$ Eq.~\eqref{polynomialbeamformer}
\EndFor
\State \textit{Select parameter estimates that maximize single-source ML approximation} $\hat{\theta}^{(i)},\hat{\cmdp}^{(i)}$ $\gets$ Eq.~\eqref{RANSAC2}, \eqref{RANSACsimple}
\State \textit{Parameter refinement based on subsampling and LS estimation} $\hat{\cmdp}^{(i)}$ $\gets$ Eq.~\eqref{refitting}\State \textit{Signal elimination in sparse support set} $\mathcal{T}^{(i+1)}$ $\gets$ Eq.~$\eqref{removecorridor}$
\EndFor
\State \textit{Parameter refinement} $\hat\cmdt^{(L)},\hat{\cmdP}^{(L)}$ \textit{based on full-source ML cost function} $f_L(\cmdt,\cmdP)$ $\gets$ Eq.~\eqref{BFGS} \\
\Return $\hat\cmdt^{(L)},\hat{\cmdP}^{(L)}$
\end{algorithmic}
\end{algorithm}
\section{Numerical Results}\label{numericalresults}
In this section, we numerically evaluate the parameter estimation performance and computational complexity of the proposed methods in Alg.~\ref{killer} and \ref{simple}, along with the methods from \cite{rakovic2017} and \cite{djurovic2019}, using Monte-Carlo simulations.

\subsection{Simulation Setup}
We consider the sonar application scenario specified in \cite{gershman2001} with a uniform linear array composed of $M=8$ omnidirectional sensors (hydrophones) with a distance $d=1.5\text{ m}$ between adjacent sensors.
The wavefront propagation speed in the medium is $c=1500 \text{ m/s}$.
Hence, the cut-off frequency for spatial aliasing is $\omega_0=\frac{\pi c}{d}=\pi \text{kHz}$, and the passband carrier frequency is $\omega_c=2\pi f_c=0.9\pi  \text{kHz}$.
The sampling period is ${\Delta=\frac{1}{f_s}=0.01 \text{s}}$ and the number of snapshots is $N=128$.
Table~\ref{table} summarizes the parameters of the $L = 4$ fourth-order PPSs under consideration, which correspond to the signal waveforms used in \cite{djurovic2018}.
The source angles in Table~\ref{table} correspond to the angular locations with maximum overlap of adjacent beamformer mainlobes at $\vartheta_{\text{MB}}(\omega_c,b\pm\frac{1}{2})$ in \eqref{DOAvsbeam}.
The receiver knows both the number of sources $L$ and the polynomial order $K = 4$.
The term $2\pi f_c$ in the initial frequency coefficient ${\phi_{\ell,1}}$ refers to the up-conversion of the baseband signal to the passband, so that the transmit signal model in \eqref{transmit} is obtained.
The baseband received signal in \eqref{received2} is obtained by down-converting the passband received signal to the baseband before sampling the signal. The complex signal amplitudes are ${\bm{\alpha}=[1,j,-1,-j]}^\mathsf{T}$ and the signal-to-noise ratio (SNR) in dB is $-10\log{\sigma^2}$, where $\sigma^2$ denotes the noise power.
 The proposed methods use $\Lambda = 500$ RANSAC iterations.
 The STFT in \eqref{y_m} employs a 64-point zero-padded DFT with a rectangular window of width $H = 16$.
 The detection threshold $\epsilon$ in \eqref{T_b} is set to the $90$th percentile of magnitude values $|z_b(p,q)|$ in \eqref{z_b}.
 An angular grid resolution of $\delta\theta = \cmddegree{1}$ is used for $\mathcal{Q}\cmdl$ in \eqref{Q} and the methods in \cite{rakovic2017} and \cite{djurovic2019}.
 The local gradient search in \eqref{BFGS} uses the BFGS Quasi-Newton algorithm with backtracking line search \cite{nocedal2006}.
 All simulations are performed on an AMD EPYC 9554P CPU.

The Root-Mean-Squared-Error (RMSE) corresponding to the estimated parameters is compared against the Cramér-Rao bound in \cite{gershman2001} that is reparametrized to match the received signal model in \eqref{stacked}.
The RMSE is defined as
\begin{equation}
	\operatorname{RMSE}(\psi)=\sqrt{\frac{1}{R L}\sum_{r=1}^{R}\sum_{\ell=1}^{L}(\psi_\ell-\cmdhat{\psi}_{r,\ell})^2}\label{rmse},
\end{equation}
where $\psi_\ell\in\left\{\theta_\ell,\phi_{\ell,1},\ldots,\phi_{\ell,K}\right\}$ denotes a true parameter of the $\ell$th source and $\cmdhat{\psi}_{r,\ell}$ is its estimate in the $r$th Monte-Carlo run.
The number of Monte-Carlo runs is $R=1000$. 
 \begin{table}[t]
	\centering
	\begin{tabular}{c||c| c |c |c| c}
		\hline
		& $\theta_\ell$ $[\text{deg}]$ & $\phi_{\ell,0}$ $[\frac{\text{rad}}{\text{s}}]$ & $\phi_{\ell,2}$ $[\frac{\text{rad}}{\text{s}^2}]$ & $\phi_{\ell,3}$ $[\frac{\text{rad}}{\text{s}^3}]$ & $\phi_{\ell,4}$ $[\frac{\text{rad}}{\text{s}^4}]$\\ [0.5ex]
		\hline
		$\ell=1$ & -25 & $2\pi(f_c-16)$ & -12 & 175 & 12\\
		$\ell=2$ & -8 & $2\pi f_c$  & 150 & 0 & 0\\
		$\ell=3$ & 8 & $2\pi f_c$  & -150 & 0 & 0 \\
		$\ell=4$ & 25 & $2\pi(f_c+16)$  & 12 & -175 & -12\\
\hline
	\end{tabular}
	\caption{Source angle and polynomial coefficients of $L=4$ PPSs under consideration \label{table}}
\end{table}
\subsection{Estimation Error Performance and Execution Time With Four PPSs}
\begin{figure}[ht!]
\centering
\begin{tikzpicture}
\begin{groupplot}[
group style ={group size =1 by 5,
vertical sep=0.5cm,
horizontal sep=1.5cm
},
width=0.45\textwidth,
height=4cm,
grid=both,
xtick={-20,-15,-10,-5,0,5,10,15,20},
ylabel style={yshift=-0.2cm},
xlabel style={yshift=0.2cm},
enlarge x limits=false,
enlarge y limits=true,
ymode=log,
grid=both,
cycle list={
{blue,mark=o},{blue,dashed,mark=square,mark options=solid},
{red,mark=o},{red,dashed,mark=square,mark options=solid},
{green,mark=o},{brown,mark=o},
{black,ultra thick},
    },
]
\nextgroupplot[thick,ylabel={RMSE($\theta$) $\left[\text{deg}\right]$},legend style={font=\small,legend columns=3,draw=none,yshift=1cm},legend to name=SNR,]
\legend{Alg.~\ref{killer}, w/o Eq.~$\eqref{BFGS}$, Alg.~\ref{simple}, w/o Eq.~$\eqref{BFGS}$, Ref.~\cite{djurovic2019}, Ref.~\cite{rakovic2017},};
\foreach \col in {1,3,4,5,6,7}{\addplot table[x index=0,y index=\col]{SNRDOA.csv};}
\addplot [black,ultra thick] table [x index=0,y index=1]{SNRCRB.csv};
\nextgroupplot[thick,ylabel={RMSE($\phi_1$) [rad/$s$]},]
\foreach \col in {1,3,4,5,6,7}{\addplot table[x index=0,y index=\col]{SNRPHI1.csv};}
\addplot  table [x index=0,y index=2]{SNRCRB.csv};
\nextgroupplot[thick,ylabel={RMSE($\phi_2$) [rad/$s^2$]},]
\foreach \col in {1,3,4,5,6,7}{\addplot table[x index=0,y index=\col]{SNRPHI2.csv};}
\addplot  table [x index=0,y index=3]{SNRCRB.csv};
\nextgroupplot[thick,ylabel={RMSE($\phi_3$) [rad/$s^3$]},]
\foreach \col in {1,3,4,5,6,7}{\addplot table[x index=0,y index=\col]{SNRPHI3.csv};}
\addplot  table [x index=0,y index=4]{SNRCRB.csv};
\nextgroupplot[grid=minor,thick,xlabel={SNR~[dB]},ylabel={RMSE($\phi_4$) [rad/$s^4$]},]
\foreach \col in {1,3,4,5,6,7}{\addplot table[x index=0,y index=\col]{SNRPHI4.csv};}
\addplot  table [x index=0,y index=5]{SNRCRB.csv};
\end{groupplot}
\path (group c1r1.north west) --node[above]{\pgfplotslegendfromname{SNR}} (group c1r1.north east);
\end{tikzpicture}
\caption{RMSE \eqref{rmse} vs. SNR for $L=4$ PPSs in Table~\ref{table} with $N=128$ snapshots and $M=8$ sensors.\label{closelyspaced}}
\end{figure}
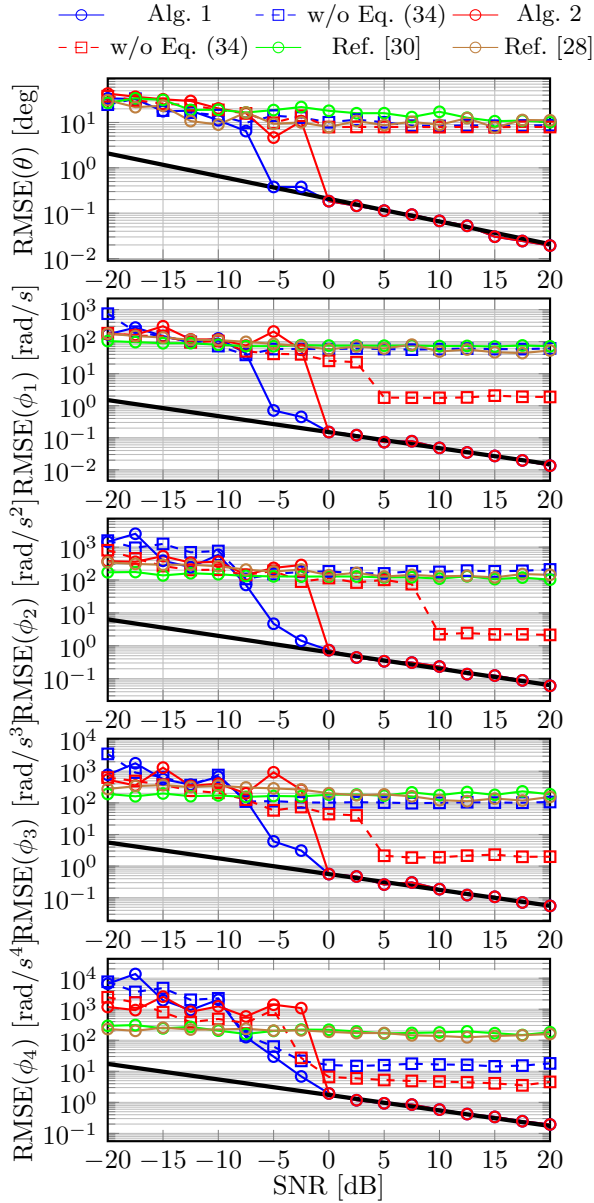

First, the estimation error performance and execution time of the methods under consideration for the $L=4$ PPSs specified in Table~\ref{table} is analyzed.
In Fig.~\ref{closelyspaced}, the RMSE of the DOA and PPS parameters is plotted for varying SNR levels.
The two proposed methods achieve the CRB for all parameters from $\text{SNR}=0~\text{dB}$. Alg.~\ref{killer} has a lower SNR threshold (about 5~dB) than Alg.~\ref{simple} for all parameters.
These results indicate that the $i$-source ML approximation yields more accurate candidate parameter estimates via RANSAC than the polynomial-phase beamformer, due to its sequentially tighter model approximation.
The local gradient search on the $i$-source ML approximation in Alg.~\ref{killer} in each sequential iteration further improves the estimation accuracy.
Results are also shown for variants of the proposed methods without performing the local gradient search in \eqref{BFGS}.
These variants exhibit significantly higher RMSE compared to the original methods, indicating that the local gradient search on the full-source ML cost function with suitable initialization is essential for approaching the CRB in multi-source DOA-PPS estimation.
The methods in \cite{rakovic2017} and \cite{djurovic2019} show high RMSE across all SNR values, possibly due to cross-terms between signal components in their cost functions and their reliance on the gradient-free Nelder-Mead optimization method \cite{nocedal2006}.
 In Fig.~\ref{runtime}, the execution time of the methods under consideration is plotted against the number of sensors $M$ at $\text{SNR}=20~\text{dB}$ to analyze the computational complexity.
It can be observed that Alg.~\ref{killer} requires about one magnitude longer execution time compared to Alg.~\ref{simple}.
This difference mainly stems from their distinct approaches to estimate the candidate DOA $\hat\theta\cmdl$ from the angular interval $\mathcal{Q}\cmdl$ in \eqref{Q}, i.e., spectral search in \eqref{RANSACtheta} versus random sampling in \eqref{random}.
For comparison, the execution time of the search-free scheme of Alg.~\ref{killer} is provided, where $\hat\theta\cmdl$ is also obtained by sampling a random angle from $\mathcal{Q}\cmdl$.
This search-free scheme of Alg.~\ref{killer} results in a considerably shorter execution time compared to the original scheme, although it is still considerably longer than that of Alg.~\ref{simple}.
The methods in \cite{rakovic2017} and \cite{djurovic2019}, which perform a spectral search over the field of view in each sequential iteration, exhibit similar execution time that is comparable to Alg.~\ref{simple}.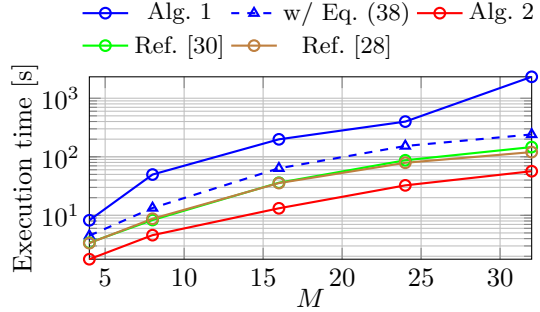
\begin{figure}[ht!]
\centering
\begin{tikzpicture}
\begin{axis}[
ymode=log,
ylabel style={yshift=-0.2cm},
grid=both,
width=0.45\textwidth,
height=4cm,
enlarge x limits=false,
enlarge y limits=false,
xlabel style={yshift=0.2cm},
xlabel={$M$},
xtick={5,10,15,20,25,30},
ylabel={Execution time [s]},
legend style={font=\small,legend columns=3,at={(0.5,1.05)},draw=none,anchor=south},
cycle list={
{blue, mark=o},{blue,dashed,mark=triangle, mark options=solid},
{red,mark=o},
{green,mark=o},{brown,mark=o},
},
] 
\legend{Alg.~\ref{killer}, w/ Eq.~\eqref{random},Alg.~\ref{simple}, Ref.~\cite{djurovic2019}, Ref.~\cite{rakovic2017} };
\foreach \col in {1,2,4,6,7}{\addplot+ [thick] table [x index=0,y index=\col]{MexecTime.csv};}
\end{axis}
\end{tikzpicture}
\caption{Execution time vs. number of sensors $M$ for $L=4$ PPSs in Table~\ref{table} with $N=128$ snapshots and $\text{SNR}=20~\text{dB}$.\label{runtime}}
\end{figure}
 \subsection{Resolution Capability for Closely Spaced Time-Frequency Spatial Source Signatures}
In this section, the capability of the methods under consideration in resolving the closely spaced time-frequency spatial signatures of two quadratic-phase signal sources is investigated.
The first source has fixed parameters $\theta_1=\cmddegree{8}$, $\phi_{1,1}=2\pi(f_c+16)$ and $\phi_{1,2}=-150\frac{\text{rad}}{s^2}$.
In each experiment, two parameters of the second source are kept identical to the first source, whereas the third parameter is varied by a positive off-set.
The SNR is $20~\text{dB}$ with $N=128$ snapshots and $M=8$ sensors.

\begin{figure}[ht!]
\centering
\begin{tikzpicture}
\begin{groupplot}[
group style ={group size =1 by 4,
vertical sep=0.6cm,
horizontal sep=1.5cm
},
width=0.45\textwidth,
height=4cm,
ylabel style={yshift=-0.2cm},
xlabel style={yshift=0.1cm},
x filter/.code={\pgfmathparse{#1*57.3}},
title style={yshift=-0.3cm},
enlarge x limits=false,
enlarge y limits=true,
xtick={0,5,10,15,20},
grid=both,
ymode=log,
cycle list={
{blue,mark=o},{red,mark=o},
{green,mark=o},{brown,mark=o},
{black,ultra thick},
},
]
\nextgroupplot[ylabel={RMSE($\theta$) [deg]},legend style={font=\small,legend columns=5,draw=none,yshift=1cm},legend to name=CommonLegend,]
\legend{Alg.~\ref{killer},Alg.~\ref{simple}, Ref.~\cite{djurovic2019}, Ref.~\cite{rakovic2017},CRB};
\foreach \col in {1,2,3,4}{\addplot+ [thick] table[x index=0,y index=\col]{deltaThetaDOA.csv};}
\addplot table [x index=0,y index=1]{deltaThetaCRB.csv};\nextgroupplot[ylabel={RMSE($\phi_{1}$) [rad/$s$]}]
\foreach \col in {1,2,3,4}{\addplot+ [thick] table[x index=0,y index=\col]{deltaThetaPHI1.csv};}
\addplot table [x index=0,y index=2]{deltaThetaCRB.csv};\nextgroupplot[xlabel={ $\delta\theta$ [deg]},ylabel={RMSE($\phi_2$) [\text{rad/$s^2$}]}]
\foreach \col in {1,2,3,4}{\addplot+ [thick] table[x index=0,y index=\col]{deltaThetaPHI2.csv};}
\addplot table [x index=0,y index=3]{deltaThetaCRB.csv};\end{groupplot}
\path (group c1r1.north west) --node[above]{\pgfplotslegendfromname{CommonLegend}} (group c1r1.north east);
\end{tikzpicture}
\vspace{-0.3cm}
\caption{RMSE \eqref{rmse} vs. angular separation $\delta\theta=\theta_2-\theta_1$ for two quadratic-phase signals with $\cmdp_1=\cmdp_2$.\label{angularres}}
\end{figure}
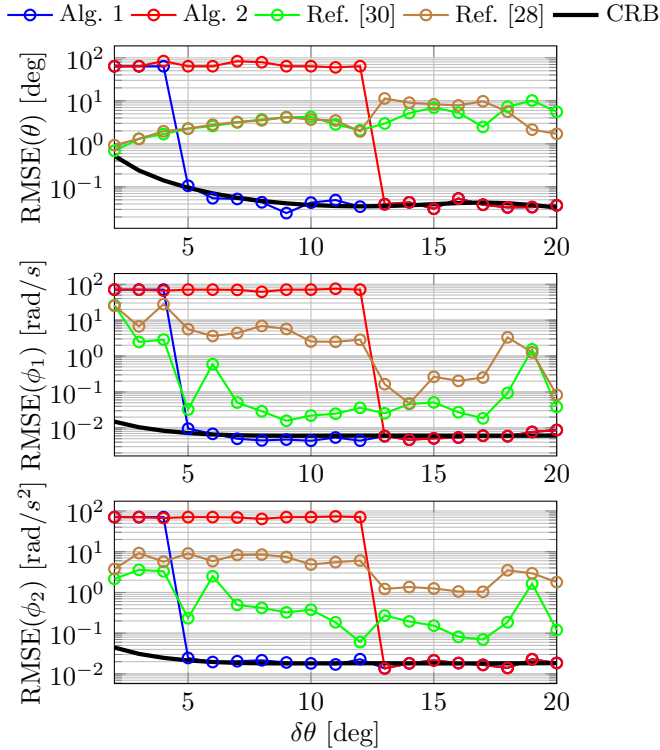

In Fig.~\ref{angularres}, the angular offset between the two sources $\delta\theta=\theta_{2}-\theta_1$ is varied between $\cmddegree{2}$ and $\cmddegree{20}$, whereas, the PPS parameters of both sources are identical, i.e., $\cmdp_1=\cmdp_2$.
It can be observed that Alg.~\ref{killer} achieves the CRB for angular separations exceeding $\delta\theta=\cmddegree{5}$, whereas Alg.~\ref{simple} has a significantly higher resolution threshold.
A cause for this high error of Alg.~\ref{simple} for low angular separations is that, in estimating the first source, the set $\mathcal{C}^{(1)}$ in \eqref{corridor0} contains the sparse support of both sources, such that, the set difference $\mathcal{T}^{(2)}=\mathcal{T}^{(1)}\setminus\mathcal{C}^{(1)}$ in \eqref{removecorridor} results in the removal of the sparse support of the second source. 
This demonstrates the limitation of the refitting step, described in \eqref{corridor0} - \eqref{refitting}, in the case of closely spaced time-frequency spatial source signatures.
On the other hand, in Alg.~\ref{killer}, the sparse support set $\mathcal{T}^{(2)}$ is obtained by processing the residual received signal in \eqref{residual} in each sequential iteration, which is also effective in the case of closely spaced source signatures.
It can be further observed that the methods in \cite{rakovic2017} and \cite{djurovic2019} struggle to resolve the sources for all $\delta\theta$ values, which is likely due to the dechirping-based extraction of estimated sources in the data that implicitly requires the source signatures to be well-separated in the time-frequency domain.
 
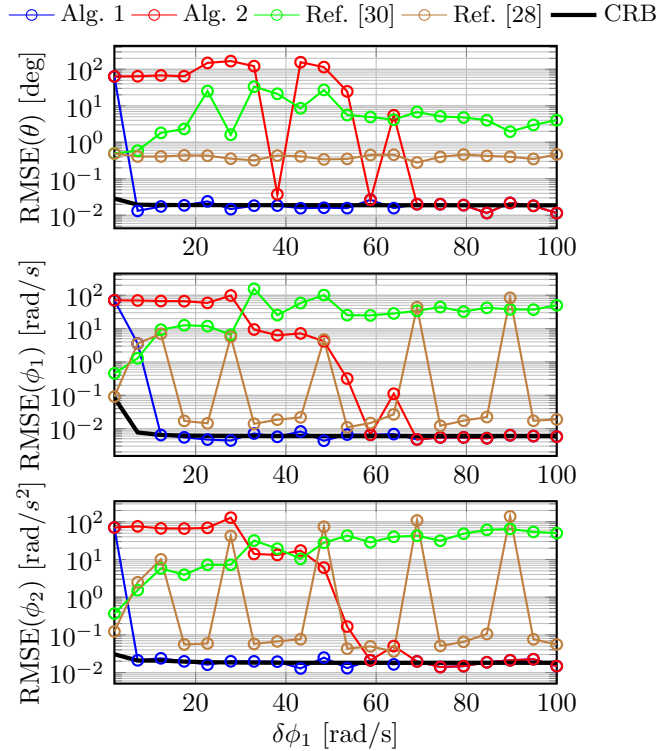
\begin{figure}[ht!]
\centering
\begin{tikzpicture}
\begin{groupplot}[
group style ={group size =1 by 4,
vertical sep=0.6cm,
horizontal sep=1.5cm
},
width=0.45\textwidth,
height=4cm,
ylabel style={yshift=-0.2cm},
xlabel style={yshift=0.1cm},
title style={yshift=-0.3cm},
enlarge x limits=false,
enlarge y limits=true,
ymode=log,
grid=both,
xtick={0,20,40,60,80,100},
cycle list={
{blue,mark=o},{red,mark=o},
{green,mark=o},{brown,mark=o},
{black,ultra thick},
},
]

\nextgroupplot[thick,ylabel={RMSE($\theta$) [deg]},legend style={font=\small,legend columns=5,draw=none,yshift=1cm},legend to name=CommonLegend,]
\legend{Alg.~\ref{killer},Alg.~\ref{simple}, Ref.~\cite{djurovic2019},Ref.~\cite{rakovic2017},CRB};
\foreach \col in {1,2,3,4}{\addplot table[x index=0,y index=\col]{deltaPhi1DOA.csv};}
\addplot table [x index=0,y index=1]{deltaPhi1CRB.csv};\nextgroupplot[thick,ylabel={RMSE($\phi_{1}$) [rad/$s$]}]
\foreach \col in {1,2,3,4}{\addplot table[x index=0,y index=\col]{deltaPhi1PHI1.csv};}
\addplot table [x index=0,y index=2]{deltaPhi1CRB.csv};\nextgroupplot[xlabel={ $\delta\phi_1$ [rad/s]},thick,ylabel={RMSE($\phi_2$) [\text{rad/$s^2$}]}]
\foreach \col in {1,2,3,4}{\addplot table[x index=0,y index=\col]{deltaPhi1PHI2.csv};}
\addplot table [x index=0,y index=3]{deltaPhi1CRB.csv};\end{groupplot}
\path (group c1r1.north west) --node[above]{\pgfplotslegendfromname{CommonLegend}} (group c1r1.north east);
\end{tikzpicture}
\vspace{-0.3cm}
\caption{RMSE \eqref{rmse} vs. frequency separation $\delta\phi_1=\phi_{2,1}-\phi_{1,1}$ for two quadratic-phase signals with $\theta_1=\theta_2$ and $\phi_{1,2}=\phi_{2,2}$. \label{phi1Res}}
\end{figure}

In Fig.~\ref{phi1Res}, the separation in the initial frequency $\delta\phi_1=\phi_{2,1}-\phi_{1,1}$ is varied between $2 \frac{\text{rad}}{\text{s}}$ and $100 \frac{\text{rad}}{\text{s}}$, whereas, the DOA and linear chirp rate of both sources are identical, i.e., $\theta_1=\theta_2$ and $\phi_{1,2}=\phi_{2,2}$.
Similarly, it can be observed that Alg.~\ref{killer} has a significantly lower resolution threshold compared to Alg.~\ref{simple}.
Precisely, the refitting step of Alg.~\ref{simple} is effective in this experiment, when the instantaneous frequency of the second source is partially outside the margins of the set $\mathcal{C}^{(1)}$ shown in Fig.~\ref{fig:corridor}, given by $\left\{\hat{\omega}^{(1)}(n)+\frac{2\pi}{\Delta H},~\hat{\omega}^{(1)}(n)-\frac{2\pi}{\Delta H}\right\}$.
Hence, the refitting step requires the difference in the instantaneous frequencies of the two signals to be larger than one frequency bin of the STFT.
This corresponds to the lower bound $\delta\phi_1=|{\omega}_2(n)-{\omega}_1(n)|>\frac{2\pi}{\Delta H}\approx39\frac{\text{rad}}{s}$ for the resolution threshold of Alg.~\ref{simple} in this experiment.
While the dechirping-based source mitigation used in the methods in \cite{rakovic2017} and \cite{djurovic2019} generally performs better when the instantaneous frequencies of the signals are better separated in frequency ($\delta\phi_1>0$), these methods still struggle to resolve the sources for all $\delta\phi_1$ values.
 
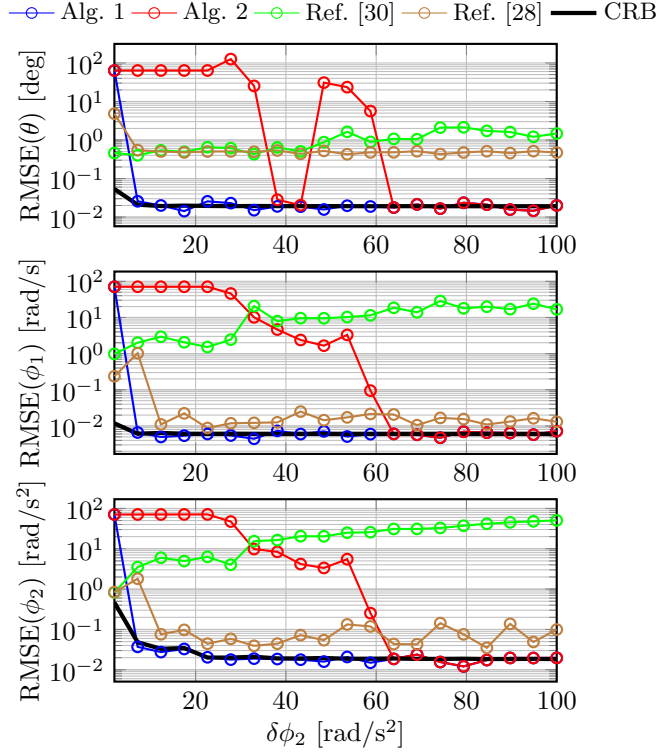
\begin{figure}[ht!]
\centering
\begin{tikzpicture}
\begin{groupplot}[
group style ={group size =1 by 4,
vertical sep=0.6cm,
horizontal sep=1.5cm
},
width=0.45\textwidth,
height=4cm,
ylabel style={yshift=-0.2cm},
xlabel style={yshift=0.1cm},
title style={yshift=-0.3cm},
enlarge x limits=false,
enlarge y limits=true,
ymode=log,
grid=both,
xtick={0,20,40,60,80,100},
cycle list={
{blue,mark=o},{red,mark=o},
{green,mark=o},{brown,mark=o},
{black,ultra thick},
},
]
\nextgroupplot[thick,ylabel={RMSE($\theta$) [deg]},legend style={font=\small,legend columns=5,draw=none,yshift=1cm},legend to name=CommonLegend,]
\legend{Alg.~\ref{killer},Alg.~\ref{simple}, Ref.~\cite{djurovic2019}, Ref.~\cite{rakovic2017},CRB};
\foreach \col in {1,2,3,4}{\addplot table[x index=0,y index=\col]{deltaPhi2DOA.csv};}
\addplot table [x index=0,y index=1]{deltaPhi2CRB.csv};\nextgroupplot[thick,ylabel={RMSE($\phi_{1}$) [rad/s]}]
\foreach \col in {1,2,3,4}{\addplot table[x index=0,y index=\col]{deltaPhi2PHI1.csv};}
\addplot table [x index=0,y index=2]{deltaPhi2CRB.csv};\nextgroupplot[xlabel={ $\delta\phi_2$ [rad/$\text{s}^2$]},thick,ylabel={RMSE($\phi_2$) [\text{rad/$\text{s}^2$}]}]
\foreach \col in {1,2,3,4}{\addplot table[x index=0,y index=\col]{deltaPhi2PHI2.csv};}
\addplot table [x index=0,y index=3]{deltaPhi2CRB.csv};\end{groupplot}
\path (group c1r1.north west) --node[above]{\pgfplotslegendfromname{CommonLegend}} (group c1r1.north east);
\end{tikzpicture}
\vspace{-0.3cm}
\caption{RMSE \eqref{rmse} vs. chirp rate separation $\delta\phi_2=\phi_{2,2}-\phi_{1,2}$ for two quadratic-phase signals with $\theta_1=\theta_2$ and $\phi_{1,1}=\phi_{2,1}$. \label{phi2Res}}
\end{figure}

In Fig.~\ref{phi2Res}, the linear chirp rate $\delta\phi_2=\phi_{2,2}-\phi_{1,2}$ is varied between $2$ and $100 \frac{\text{rad}}{\text{s}^2}$, whereas, the DOA and initial frequency of both sources are identical, i.e., $\theta_1=\theta_2$ and $\phi_{1,1}=\phi_{2,1}$. 
Similarly, it can be observed that Alg.~\ref{killer} has a significantly lower resolution threshold compared to Alg.~\ref{simple}.
In this experiment, the time-dependent difference in the instantaneous frequencies of the signals is given by $\delta\omega(n)=\omega_2(n)-\omega_1(n)=2(\delta\phi_2)(\Delta n)$.
The refitting step of Alg.~\ref{simple} is thus effective in this experiment, when the instantaneous frequency of the second source is partially outside the margins $\left\{\hat{\omega}^{(1)}(n)+\frac{2\pi}{\Delta H},~\hat{\omega}^{(1)}(n)-\frac{2\pi}{\Delta H}\right\}$ in Fig.~\ref{fig:corridor}, i.e., $\max\limits_{n}|\delta\omega(n)|>\frac{2\pi}{\Delta H}$, which corresponds to the lower bound $\delta\phi_2>\frac{2\pi}{\Delta^2 (N-1)H}\approx 61\frac{\text{rad}}{s^2}$ for the resolution threshold of Alg.~\ref{simple} in this experiment. Although the dechirping-based source mitigation in \cite{rakovic2017} and \cite{djurovic2019} performs better with larger frequency separation between signals ($\delta\phi_2>0$), these methods struggle to resolve the sources across all tested $\delta\phi_2$ values.
 
\section{Conclusion}\label{conclusion}
In this work, sequential estimators for the multi-source DOA-PPS estimation problem have been proposed.
The measurement data is transformed into the time-frequency-beamspace domain through sensor-wise time-frequency analysis and coherent array processing.
The algorithm leverages the sources' sparse representation and employs RANSAC to estimate the DOA and polynomial-phase parameters of one source at a time using sequential approximations of the ML cost function.
Numerical experiments demonstrate that the proposed methods can achieve the Cramér-Rao bound of all parameters.
The low-cost variant of the proposed methods is computationally more efficient, but shows limitations in resolving sources with closely spaced time-frequency spatial signatures.

In future work, improving the accuracy of the source parameter estimates obtained from the RANSAC method in \eqref{RANSAC2} is of high interest, because it can be observed in Fig.~\ref{closelyspaced} that the RMSE of the proposed methods without performing the local gradient search is poor.
This can be accomplished, e.g., through time-variant beamforming that is matched to the change in the instantaneous frequency of the impinging PPSs to achieve fully coherent processing of the array measurements.
Similarly, subarray processing can also be considered, where the size of the subarrays is defined by the space-time interval described in \eqref{approximation} for coherent processing of the measurements.
While this work focuses on sequential estimation of individual sources, future research could explore simultaneous estimation of source pairs.
As discussed in Section \ref{sparsesupport}, estimating all sources simultaneously with RANSAC leads to prohibitive computational complexity.
However, a modified sequential approach that estimates two sources per iteration could potentially maintain estimation accuracy while reducing computational cost compared to the proposed methods.

\appendix
\section{Efficient $i$-Source ML Approximation} \label{appendix:isource}
Following a similar approach as in \cite{trinh-hoang2018}, we derive the efficient implementation of the $i$-source ML approximation in \eqref{isource:simplified}, which is obtained from the original $i$-source ML approximation in \eqref{isource}, by assuming that the source parameter estimates in the previous $(i-1)$ sequential iterations are fixed.
For simplicity of notation, we denote the array response matrix in the $\lambda$th RANSAC iteration of the $i$th iterations as $[\Ab,\bar{\bm{a}}]\in\Cbb^{MN\times i}$, where $\Ab=\Ab(\hat{\cmdt}^{(i-1)},\hat{\cmdP}^{(i-1)})\in\Cbb^{MN\times (i-1)}$ is the array response matrix of the $(i-1)$ estimated sources in the previous iterations and $\bar{\bm{a}}=\ab(\vartheta,\hat{\cmdp}\cmdl)\in\Cbb^{MN}$ is the array response vector of the $i$th source with source angle $\vartheta$ and PPS parameter estimate $\hat{\cmdp}\cmdl$ from \eqref{polynomialregression}.

The original $i$-source ML approximation in \eqref{isource} can then be expressed as
\begin{equation}\label{isource2}
	f_i(\vartheta,\hat{\cmdp}\cmdl) = \|\bm{\Pi}^\perp_{[\Ab,\ab]}\cmdxbar\|_2^2.
\end{equation}
Using the property $\cmdorthPi_{[\Ab,\ab]}=\cmdorthPi_{\Ab}-\cmdPi_{\cmdorthPi_{\Ab}\ab}$, this cost function can be rewritten as 
\begin{equation}\label{isource:OMP}
		f_i(\vartheta,\hat{\cmdp}\cmdl)=|\cmdxbar\herm\cmdorthPi_{\Ab}\cmdxbar-2\cmdxbar\herm(\cmdorthPi_{\Ab})\herm\cmdPi_{\at}\cmdxbar+\cmdxbar\herm\cmdPi_{\at}\cmdxbar|,
\end{equation}
where $\at=\cmdorthPi_{\Ab}\ab$ and $\cmdPi_{\at}=\frac{\at\at\herm}{\|\at\|_2^2}$.
The efficient $i$-source ML approximation is then obtained as
\begin{equation}
f_i(\vartheta,\hat{\cmdp}\cmdl)=\|\cmdr\cmdim\|_2^2-\frac{|\ab\herm\cmdr\cmdim|^2}{\|\at\|_2^2},
\end{equation}
where $\cmdr\cmdim=\cmdorthPi_{\Ab}\cmdxbar$ is the residual received signal in \eqref{residual}. 
This efficient implementation is related to the orthogonal matching pursuit (OMP) method, which has recently been applied in DOA estimation \cite{tropp2007,trinh-hoang2020b,pesavento2023}. 
\section{Gradient Derivation \label{appendix:gradient}}
For the local gradient search in the $i$th sequential iteration in \eqref{BFGS}, the gradient of the $i$-source ML approximation $f_i(\cmdt,\cmdP)$ in \eqref{isource} with respect to the DOAs $\cmdt\in\Rbb^i$ and PPS parameters $\cmdP\in\Rbb^{K\times i}$ of the $i$ sources is required.
For simplicity of notation, the array response matrix is denoted  as $\cmdAbarpsi=\cmdAbar\in\Cbb^{MN\times i}$.
 The $i$-source ML approximation in \eqref{isource} can be expressed as
\begin{align}
\begin{split}
f_i(\cmdt,\cmdP)&=\big\|\bm{\Pi}^\perp_{\cmdAbarpsi}\cmdxbar\big\|_2^2,\\
	&=|\cmdxbar\herm\cmdxbar-\cmdxbar\herm\cmdAbarpsi\big(\cmdAbarpsi\herm\cmdAbarpsi\big)^{-1}\cmdAbarpsi\herm\cmdxbar|.
\end{split}
\end{align}
The differential $df_i(\cmdt,\cmdP)$ in dependence of the differential $d\cmdAbarpsi$ can then be obtained using the product rule \cite{hjorungnes2011}, given by
\begin{equation}
df_i(\cmdt,\cmdP)=-\Big|\cmdxbar\herm (d\cmdAbarpsi)\big(\cmdAbarpsi\herm\cmdAbarpsi\big)^{-1}\cmdAbarpsi\herm\cmdxbar+\cmdxbar\herm\cmdAbarpsi (d\big(\cmdAbarpsi\herm\cmdAbarpsi\big)^{-1})\cmdAbarpsi\herm\cmdxbar+\cmdxbar\herm\cmdAbarpsi\big(\cmdAbarpsi\herm\cmdAbarpsi\big)^{-1}(d\cmdAbarpsi\herm)\cmdxbar\Big|,
\end{equation}
Given that $\cmdAbarpsi\herm\cmdAbarpsi\in\Cbb^{i\times i}$ is a complex-valued square matrix $\bm{Z}$, the differential properties ${d\bm{Z}^{-1}=-\bm{Z}^{-1}(d\bm{Z}) \bm{Z}^{-1}}$ and $d(\bm{Z}^H)=(d\bm{Z})^H$ \cite{hjorungnes2011} can be used to simplify the differential $df_i(\cmdt,\cmdP)$ to
\begin{equation}
	df_i(\cmdt,\cmdP)=
	\Big|\hat{\bm{\alpha}}\herm d\Big( \cmdAbarpsi\herm\cmdAbarpsi\Big)\hat{\bm{\alpha}}\Big|
	-2\operatorname{Re}\Big\{\bm{x}\herm (d\cmdAbarpsi)\hat{\bm{\alpha}}
	\Big\},
\end{equation}
where the amplitude vector $\hat{\bm{\alpha}}=\cmdAbarpsi^\dagger\cmdxbar\in\Cbb^i$ is obtained according to \eqref{concentration}.
 The left term can be rewritten using the product rule $d(\bm{Z}_1\bm{Z}_2)=(d\bm{Z}_1)\bm{Z}_2+\bm{Z}_1(d\bm{Z}_2)$. The differential $df_i(\cmdt,\cmdP)$ can then be summarized as
\begin{equation}
	df_i(\cmdt,\cmdP)=2\operatorname{Re}\Big\{
	\hat{\bm{\alpha}}\herm\cmdAbarpsi\herm (d\cmdAbarpsi)\hat{\bm{\alpha}}
	-\bm{x}\herm (d\cmdAbarpsi)\hat{\bm{\alpha}}
	\Big\}.\label{dfi}
\end{equation}
The derivative of $\cmdAbarpsi$ with respect to the scalar parameter $\psi\in\{\theta_1,\phi_{1,1},\ldots,\phi_{1,K},\ldots,\theta_i,\phi_{i,1},\ldots,\phi_{i,K}\}$ is given by 
 \begin{equation}
	 \bar{\bm{D}}=\frac{{d}\cmdAbarpsi}{d\psi}\in\Cbb^{MN\times i}.
 \end{equation}
The gradient of the $i$-source ML approximation with respect to $\psi$ is then obtained as\begin{equation}
 		\frac{df_i(\cmdt,\cmdP)}{d\psi}=2\operatorname{Re}\Big\{\hat{\bm{\alpha}}\herm\Big(\cmdAbarpsi\herm\bar{\bm{D}}\Big)\hat{\bm{\alpha}}-\cmdxbar\herm\bar{\bm{D}}\hat{\bm{\alpha}}\Big\}.
 \end{equation}
 \section*{Acknowledgment}
This work has received funding from the German Federal Ministry of Education and Research (BMBF) within the project “Open6GHub” under grant number 16KISK014.
\bibliographystyle{elsarticle-num}
\bibliography{References}
\end{document}